\input{psfig.sty}
\input{epsf.sty}

\def\mev{\,{\rm Me\kern-0.1em V}}
\def\gev{\,{\rm Ge\kern-0.1em V}}

\documentstyle[12pt]{article}

\renewcommand{\baselinestretch}{1.7}
\begin{document}
\vspace*{-1.25in}
\small{
\begin{flushright}
FERMILAB-Pub-00/154-T \\[-.1in] 
July~2000 \\
\end{flushright}}
\vspace*{.75in}
\begin{center}
{\Large{\bf  Anomalous Chiral Behavior\\ in Quenched Lattice QCD}}\\
\vspace*{.45in}
{\large{W. ~Bardeen$^1$,
A.~Duncan$^2$, 
E.~Eichten$^1$,  and
H.~Thacker$^3$}} \\ 
\vspace*{.15in}
$^1$Fermilab, P.O. Box 500, Batavia, IL 60510 \\
$^2$Dept. of Physics and Astronomy, Univ. of Pittsburgh, 
Pittsburgh, PA 15260\\
$^3$Dept.of Physics, University of Virginia, Charlottesville, 
VA 22901
\end{center}
\vspace*{.3in}
\begin{abstract}
A study of the chiral behavior of pseudoscalar masses and decay constants is
carried out in quenched lattice QCD with Wilson fermions. Using the modified
quenched approximation (MQA) to cure the exceptional configuration problem,
accurate results are obtained for pion masses as low as $\approx$ 200 MeV. The 
anomalous chiral log effect
associated with quenched $\eta'$ loops is studied in both the relation between $m_{\pi}^2$
vs. $m_q$ and in the light-mass behavior of the pseudoscalar and axial vector 
matrix elements. The 
size of these effects agrees quantitatively with a direct measurement of the
$\eta'$ hairpin graph, as well as with a measurement of the topological susceptibility,
thus providing several independent and quantitatively consistent determinations of the 
quenched chiral log parameter $\delta$. For $\beta=5.7$ with clover-improved 
fermions $(C_{sw}
=1.57)$ all results are consistent with $\delta=0.065\pm 0.013$ .
\end{abstract}

\section{Introduction}

Until recently, efforts to study the chiral limit of lattice QCD with Wilson-Dirac
fermions have been impeded by the large non-gaussian fluctuations  encountered
at light quark mass (the ``exceptional configuration'' problem). These statistical
problems have been shown to be symptomatic of the nonconvergence of the quenched lattice 
Monte Carlo integration due to the presence of exactly real 
Dirac eigenmodes in the physical mass region \cite{MQA2}.
These real eigenmodes are of topological origin but are displaced from zero mass by
the explicit chiral symmetry breaking of the Wilson-Dirac operator.
The modified quenched approximation (MQA) is a prescription for constructing a
modified quark propagator by shifting positive mass real poles to zero mass in a
compensated way.\cite{MQA} This procedure is the most
straightforward and effective way of removing the lattice artifact which produces exceptional
configurations while leaving the results
for heavier quarks outside the pole region essentially unchanged. In the continuum limit,
the spread of eigenmodes to positive mass  shrinks to zero, so that if
we shift all poles above a given quark mass, the fraction of gauge configurations in an ensemble
that would require MQA pole-shifting approaches zero in the continuum limit.
In this sense, the MQA has the right continuum limit, and 
provides a sensible definition of the otherwise
undefined quenched theory at finite lattice spacing and small quark mass.
Exactly how effective the MQA is in restoring the chiral behavior
expected from quenched continuum arguments is a question that can best be addressed
by detailed calculations.
As we will show in this paper, by using the MQA procedure we are able to
study the behavior of pseudoscalar masses and decay constants with high
statistics down to much lower pion mass (ca. $200$ MeV) than has previously been accessible
at $\beta=5.7$ (for example, a recent high statistics study \cite{CPPACS} of quenched chiral behavior
extends only to $m_{\pi} \approx 300$ MeV). The results confirm in considerable detail the expected chiral 
behavior predicted for these quantities from chiral Lagrangian 
arguments. The overall size of the quenched chiral log parameter $\delta$ is
about a factor of two smaller than the $\delta\approx 0.17$ 
expected from the theoretical result
\begin{equation}
\label{eq:delta}
\delta = \frac{m_0^2}{16\pi^2f_{\pi}^2 N_{f}}\;,
\end{equation}
using the physical values of
$m_{\eta'}$ and $f_{\pi}$. (Here the normalization of $f_{\pi}$ is such that its physical
value is $\approx 95$ MeV. Chiral perturbation theory and the physical mass of the 
$\eta'$ give $m_0\approx 850$ MeV.) 
Nevertheless, the quantitative relations between various quantities
inferred from continuum chiral Lagrangian arguments appear to be well-satisfied.
In fact, the smallness of $\delta$ is consistent with our previous calculations at heavier
quark mass which did not use MQA shifting \cite{lat96}.

In this paper, we describe several independent quantitative estimates of the
quenched chiral log (QCL) parameter $\delta$: (1) a direct measurement of the $\eta'$
hairpin diagram, (2) a calculation of the topological susceptibility combined with
the Witten-Veneziano formula \cite{WitVen}, (3) measurement of the QCL effect
in $m_{\pi}^2$  vs. $m_q$ and
in the vacuum-to-one-particle matrix element of the pseudoscalar density,
(4) by fitting cross-ratios of masses and matrix elements using the lowest
order  chiral perturbation theory  prediction (to first order in $\delta$),
 and (5) by a global fit of all masses and matrix elements to the next-to-leading
 order chiral perturbation theory prediction.
When the two quark masses in the pion are allowed to vary independently, as
in (4) and (5), 
the corresponding masses and decay constants provide additional tests of quenched
chiral log structure (c.f. the analysis of the CPPACS collaboration\cite{CPPACS}).
In this paper we present the results of analyses of chiral log structure both in the
``diagonal" (equal quark mass) sector, and for mesons with unequal quark masses.
The various determinations of  
$\delta$ are all consistent within statistics, giving an exponent $\delta=.065(13)$
for clover-improved Wilson fermion action, and $\delta=.06(2)$ for unimproved 
Wilson.
 
As discussed below, operators which do not change chirality, such as the
axial-vector current, are not expected to exhibit 
any QCL singularity (for equal mass quarks).
In agreement with this theoretical expectation,
our results for the mass dependence of the axial-vector decay constant show smooth
analytic behavior with no enhancement in the chiral limit. The contrast between the
singular chiral behavior of the pseudoscalar matrix element and the smooth analytic 
behavior of
the axial-vector matrix element is
a particularly convincing piece of evidence that we are indeed seeing
the effects of virtual $\eta'$ loops. 

 In Section 2 we briefly review the essential features of quenched chiral behavior in
QCD in the language of effective field theory.
 Details of the lattice calculations are
presented in Section 3. The direct determinations of $\delta$ from the $\eta'$ 
hairpin correlator and from the topological susceptibility are presented in Section 4. 
In Section 5, QCL effects in the squared pseudoscalar mass as a function of quark mass
are discussed, for ``diagonal" mesons composed of equal mass quarks only. 
The QCL behavior of the pseudoscalar and axial-vector decay constants(again, for diagonal
mesons only) is studied in Section 6. In Section 7 we extract the chiral log
parameter $\delta$ from an analysis of cross-ratios of pseudoscalar meson masses
 and decay constants, along the lines of the CPPACS analysis \cite{CPPACS}.
In Section 8 we compare our lattice data  for all
 mesons (equal and unequal quark masses) with the
results of a detailed quenched chiral perturbation theory (to order $p^4$)
 calculation for the pseudoscalar
masses and decay constants. 
Finally, in Section 9 we summarize our results and draw some broad conclusions. Some
technical remarks on the efficacy and accuracy of the all-source technique used to
 extract hairpin amplitudes are relegated to Appendix 1, while chiral perturbation theory
 formulas for pseudoscalar and axial decay constants are given in Appendix 2.

\newpage
\section{Brief review of quenched chiral theory}

\begin{figure}
\vspace*{2.0cm}
\includegraphics{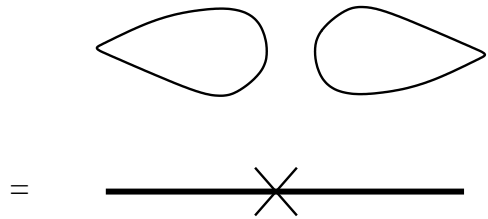}
\vspace{3.5cm}
\caption[]{Chiral Lagrangian representation of the quark hairpin diagram as an $\eta'$
mass insertion.  }
\label{fig:hairpin}
\end{figure}

\begin{figure}
\vspace*{2.0cm}
\includegraphics{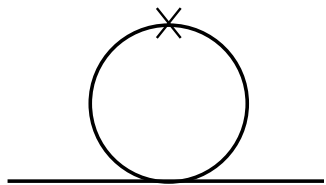}
\vspace{3.5cm}
\caption[]{A quenched $\eta'$ loop diagram which produces anomalous chiral log behavior.}
\label{fig:etaloop}
\end{figure}

The presence of anomalous chiral behavior induced by the quenched approximation
was first  investigated in the work of Sharpe \cite{Sharpe} and Bernard and Goltermann
\cite{B&G}. Physically, the quenched chiral log effect can be understood as
a consequence of the absence of topological screening in quenched QCD. Diagramatically,
the quenched approximation discards some, but not all, of the $\eta'$ hairpin 
mass insertions. Instead of the massive $\eta'$ propagator of full QCD, the
quenched hairpin diagram exhibits a double Goldstone pole, see Fig[\ref{fig:hairpin}].
This leads to infrared singular $\eta'$ loop diagrams, such as that shown in Fig[\ref{fig:etaloop}].
which alter the chiral behavior of certain matrix elements. A convenient way to understand
the effect of the QCL singularities is to interpret them as a renormalization
factor in the chiral field. Begin with a $U(3)\times U(3)$ chiral Lagrangian with a
chiral field
\begin{equation}
U = \exp\left[i\sum_{i=0}^8\lambda_i\pi_i/f \right]\equiv e^{i\pi_0/f}\tilde{U}
\end{equation}
where $\lambda_0=1$ and $\pi_0\equiv\eta'$ represents the SU(3)-flavor singlet meson.
Now consider the effect of integrating out the $\eta'$ field.
The remaining $SU(3)\times SU(3)$ Goldstone fields will be renormalized

\begin{equation}
U\rightarrow \langle e^{i\pi_0/f}\rangle\tilde{U}=\exp\left[-\langle
\pi_0^2\rangle/2f^2\right]\tilde{U}
\end{equation}
where, in the quenched theory, 
\begin{equation}
 <\pi_{0}^{2}> = \int\frac{d^{4}p}{(2\pi)^{4}}\frac{1}{p^{2}+M_{\eta^{\prime}}^{2}}
\rightarrow \int\frac{d^{4}p}{(2\pi)^{4}}\frac{-m_{0}^{2}}{(p^{2}+M_{\pi}^{2})^{2}}
\end{equation}

In full QCD, the $\eta'$ mass is large, so the effect of integrating out the $\eta'$
field is to induce a finite renormalization that is nonsingular in the chiral limit.
The chiral limit is described by the $SU(3)\times SU(3)$ chiral Lagrangian
without an $\eta'$. Now consider the effect of quenching on the result of
integrating out the $\eta'$ field. The double Goldstone pole in the quenched
$\eta'$ propagator produces logarithmically divergent loop integrals in
chiral perturbation theory. This results in a renormalization factor for the field $U$
that is singular in the chiral limit, and thus alters the $m_{\pi}\rightarrow 0$ 
behavior of the chiral field:
\begin{equation}
U\rightarrow \exp\left[-\delta\log m_{\pi}^2\right]\tilde{U} =\left(\frac{1}{m_{\pi}^2}\right)^{\delta}
\tilde{U}
\end{equation}
where $\tilde{U}$ is nonsingular as $m_{\pi}\rightarrow 0$.
In the following subsections, we will discuss the chiral log effect (or lack
thereof) in the pseudoscalar and axial-vector charge matrix elements and in the
mass of the pion as a function of quark mass.

  Define the decay constants
$f_P$ and $f_A$ 
\begin{eqnarray}
\langle 0|\bar{\psi}\gamma^5\psi|\pi(p)\rangle & = & f_P \nonumber\\
\langle 0|\bar{\psi}\gamma^5\gamma^{\mu}\psi|\pi(p)\rangle & = & p^{\mu}f_A
\end{eqnarray}
From the chiral-field expressions for the quark bilinears,
\begin{eqnarray}
\bar{\psi}\gamma^5\psi & \propto & U-U^{\dag} \nonumber \\
\bar{\psi}\gamma^5\gamma^{\mu}\psi & \propto & i\left(U^{-1}\partial^{\mu}U
-(\partial^{\mu}U^{-1})U\right)\;\;,
\end{eqnarray}
we conclude that a singular chiral log factor will appear in the quenched
calculation of $f_P$, but not of $f_A$,
\begin{eqnarray}
f_P^{\rm quenched} &=& \left(\frac{1}{m_{\pi}^2}\right)^{\delta}\tilde{f}_P \nonumber \\
f_A^{\rm quenched} &=& \tilde{f}_A
\end{eqnarray}
where $\tilde{f}_P$ and $\tilde{f}_A$ go to a constant in the chiral limit.
Both of these expectations are confirmed by the data, as discussed below. 
The chiral behavior of the pion mass as a function of quark mass is easily derived
from the results for the pseudoscalar and axial-vector densities, combined
with PCAC. This gives
\begin{equation}
m_{\pi}^2 \simeq {\rm const.}\times m_q \times (\frac{1}{m_{\pi}^{2}})^{\delta},\;\;\;m_q \rightarrow 0
\end{equation}
This predicted behavior is also confirmed by the lattice results, as discussed in 
Section 5.

The quantitative significance of the quenched chiral log behavior that we observe is further
reinforced by comparing it with a direct calculation of the anomalous exponent
$\delta$. We have done this calculation in two related but distinct ways.
One is a calculation of the hairpin diagram Fig[\ref{fig:hairpin}].
\begin{equation}
\label{eq:hairpin}
\Delta_h(x) = \langle{\rm Tr}\gamma^5G(x,x){\rm Tr}\gamma^5G(0,0)\rangle
\end{equation}
where $G(x,y)$ is the quark propagator. 
The size of the hairpin mass insertion determines 
the coefficient of the logarithmic divergence in the $\eta'$ loop diagrams. 
Let us denote by $m_0^2$ the value of the $\eta'$ mass insertion vertex obtained from
the long-range behavior of the hairpin diagram. Then the predicted chiral log
exponent is given by Equation (\ref{eq:delta}).

The other quantity which provides
an evaluation of $\delta$ is the topological susceptibility, $\chi_t$.
The Witten-Veneziano formula allows us to relate $\chi_t$ to the 
$\eta'$ mass insertion,
\begin{equation}
\label{eq:WV}
m_0^2 = \frac{2N_f}{f_{\pi}^2}\chi_t
\end{equation}
where $N_f$ is the number of (light) quark flavors.
The topological susceptibility is given in terms of winding numbers $\nu$ on a lattice
of four-volume $V$ by
\begin{equation}
\chi_t = \frac{\langle\nu^2\rangle}{V}
\end{equation}
To calculate winding numbers, we use the integrated anomaly technique (``fermionic method'')
of Smit and Vink \cite{SmitVink}. 
The configuration-by-configuration calculation of winding numbers provides a determination
of the $\chi_t$, and hence via (\ref{eq:WV}) and (\ref{eq:delta}), an estimate of $\delta$.
The results obtained from the hairpin residue and from the topological susceptibility
are in good agreement for clover improved fermions, as discussed in Section 4. 

\newpage
\section{Lattice calculation of pseudoscalar masses and decay constants}

Most of the results which we will focus on in the subsequent discussion are obtained
from the ``b'' ensemble of gauge configurations from
the Fermilab ACPMAPS library. This ensemble consists of a set of 300
quenched configurations on a $12^3\times 24$ lattice at $\beta=5.7$.
The fermion action used in our calculations was clover improved with a clover coefficient
of $C_{sw}=1.57$. 
To get some estimate of lattice spacing effects and finite volume corrections, we also 
calculated masses and decay constants on 200 configurations of the b-lattice
ensemble with unimproved Wilson fermions ($C_{sw}=0$), and on the a-lattice ensemble
($16^3\times 32$ at $\beta=5.7$), also with unimproved Wilson fermions.
The masses and decay constants and their quoted errors are obtained from fully correlated
fits to the smeared-smeared and smeared-local propagators, using a smeared
pseudoscalar source in Coulomb gauge combined with local pseudoscalar and axial-vector sources.
Calculations with different smearing 
functions give results that are consistent
within less than one standard deviation, indicating that the systematic error
associated with excited state contamination is less than our statistical
errors. For the hairpin correlator, the issue of excited state contamination
is addressed in detail in Section 3.2 by comparing the ratio of smeared and
local hairpin propagators. Remarkably, we find that for small quark masses there
is very little excited state contribution to the hairpin propagator, even at
time separations as small as $t=1$. This is in marked contrast to the valence pion
propagator, which has substantial excited state components at short time.
We conclude that the hairpin vertex itself is largely 
decoupled from the excited states in the pseudoscalar channel.

The essential new ingredient in the present study which allows a detailed exploration
of chiral behavior is the use of the pole-shifting procedure of the modified 
quenched approximation. The details of this procedure and its effectiveness in resolving
the exceptional configuration problem have been described previously \cite{MQA}.
For each gauge ensemble and choice of fermion action, we carried out a careful
scan of each configuration for poles over a range of quark masses, starting at a
 value  heavy enough
to be beyond all real eigenmode poles and going to nearly zero mass.
For example, in the run with clover action, where $\kappa_c=0.14329$, poles appeared
as low as $\kappa\approx 0.1417$ ($m_q \simeq$ 45 MeV),  and we scanned for and located
all poles up to 0.1431 . 
The value of
the integrated pseudoscalar charge $Q_5=\int d^4x\bar{\psi}\gamma^5\psi$ is calculated
for a sequence of hopping parameter values, using the same allsource method that
is used to calculate the hairpin propagator (see below). A Pade fit to these $Q_5$
values determines the location of any poles within and somewhat beyond the range
scanned. Extremely precise pole locations can be obtained by 
performing further conjugate gradient inversions very close to the pole positions
determined by the Pade fit. Using the stabilized biconjugate gradient algorithm, we are able to perform
inversions at hopping parameters very close to the pole position without any
major increase in convergence time. 
In our calculations, we have located the pole positions as a
function of hopping parameter $\kappa$ to at least 8-digit accuracy in all
cases. Once all the visible poles in an ensemble are located, their residues
in the quark propagator are determined by performing inversions slightly above
and below the pole and subtracting. We found that accurate pole residues
could be computed for a pole at $\kappa_0$ by inverting at $\kappa=\kappa_0
\pm .000001$. (The computation of pole residues can be done very economically
by noting that the pole contribution to all 12 color-spin components of the
quark propagator can be obtained from a single color-spin inversion above and
below the pole, i.e. only $1/12$ of a full propagator calculation is required.) 

\begin{figure}
\vspace*{2.0cm}
\includegraphics{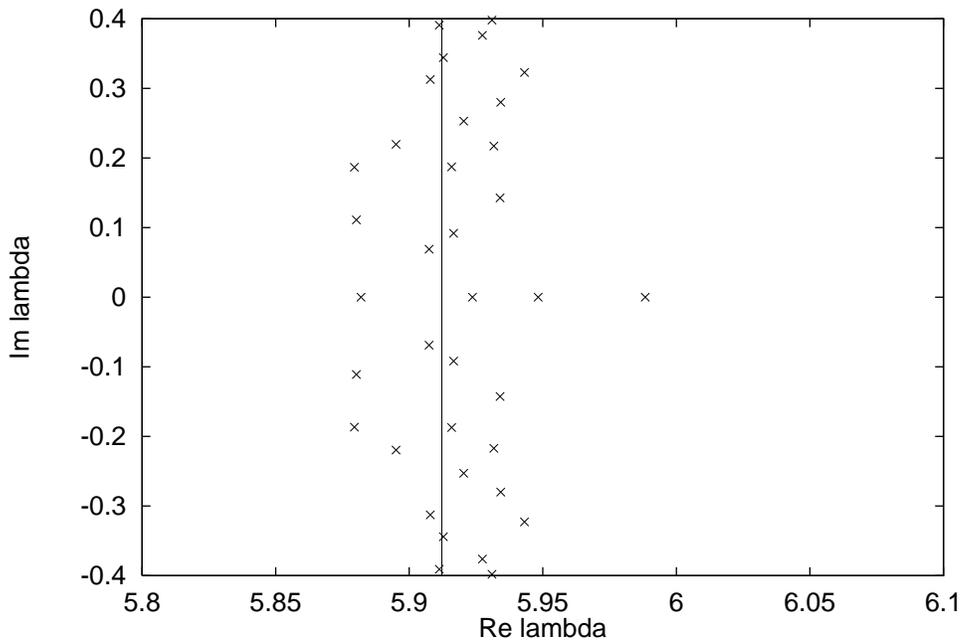}
\vspace{6.5cm}
\caption[]{A portion of the low-lying spectrum of the $C_{sw}=0$ Wilson-Dirac spectrum
of a  quenched gauge configuration (b\_021000 of 
the ACPMAPS library) in the region $\lambda\approx 1/\kappa_c$. This configuration has four 
real poles in the continuum band. The vertical line is at $Re\;\lambda=1/\kappa_c$. 
The three real poles farthest to the right on the plot require shifting.}
\label{fig:lowspec}
\end{figure}

An alternative procedure for locating poles of the quark propagator
$\propto 1/(D\!\!\!/-m_q)$ is to use the Arnoldi algorithm \cite{Arnoldi} to partially
diagonalize the Wilson-Dirac operator in the region around zero mass. The
returned eigenvalues are the pole positions, and the residues needed to
perform the pole-shifting procedure may be reconstructed from the Arnoldi eigenvectors.
This method locates not only the real eigenvalues but also the complex
ones in the continuum band. The Arnoldi analysis has been carried out \cite{McCune}
on a subset of the configurations used in this
investigation, extracting approximately
50 eigenvalues from each gauge configuration analyzed. For the lattices which 
exhibited visible poles in the bicongradstab scanning procedure described
above, the results of the Arnoldi calculation agree accurately with 
that analysis for both the pole positions and residues. The low-lying
spectrum for an ``exceptional''  gauge configuration ($b:021000$) is shown in Fig[\ref{fig:lowspec}].
(Note: The positive mass region is to the right in this plot. The three modes farthest
to the right on the real axis were MQA shifted.)
  
The effect of the MQA pole-shifting procedure is to eliminate the
problem of exceptional configurations and to dramatically improve error bars
on all quantities calculated from light quark propagators, as shown in detail
in Ref.\cite{MQA}. Since the $\eta'$ hairpin propagator is particularly sensitive
to the topological structure of gauge configurations, the improvement obtained
by using the MQA for the hairpin calculations is even more striking than that for
valence-quark meson propagators. The MQA-improved results are sufficiently accurate 
to allow a detailed study of the time-dependence
of the hairpin propagator even as far out as $t = 9$ or 10. Over the entire time range
starting from $t=1$,
this time-dependence is in reasonably good agreement
with that predicted from a pure double pion pole form. The lack of any significant 
single-pole contribution to the hairpin propagator is consistent with the absence of
excited states. (A heavy excited state on one side of the hairpin vertex would produce an
effective single pion pole term.)  A single-pole 
term in the time-dependence could also arise from a 
$p^2$-dependent hairpin vertex. In the chiral Lagrangian, this would correspond
to a renormalization of the $\eta'$ kinetic term in addition to a mass insertion.
The fact that the time-dependence of the hairpin
correlator is well-described by the pure double pion pole form over the
entire time range from $t=1$ to $12$ lends quantitative support
to the assumption, often used in phenomenological discussions, that the
momentum dependence is mild and the  $\eta'$
hairpin can therefore be treated simply as a mass insertion. The time-dependence is discussed in
more detail in Section 4.2. The final results quoted for the mass insertion value $m_0^2$ are
obtained from a 1-parameter fit of the hairpin to a pure double-pole formula,
with the pion mass held fixed at the value obtained from the corresponding
valence pion propagator.

\newpage
\section{The $\eta'$ mass and the chiral log parameter $\delta$}

\subsection{Topological susceptibility}

\begin{figure}
\vspace*{0.5cm}
\includegraphics{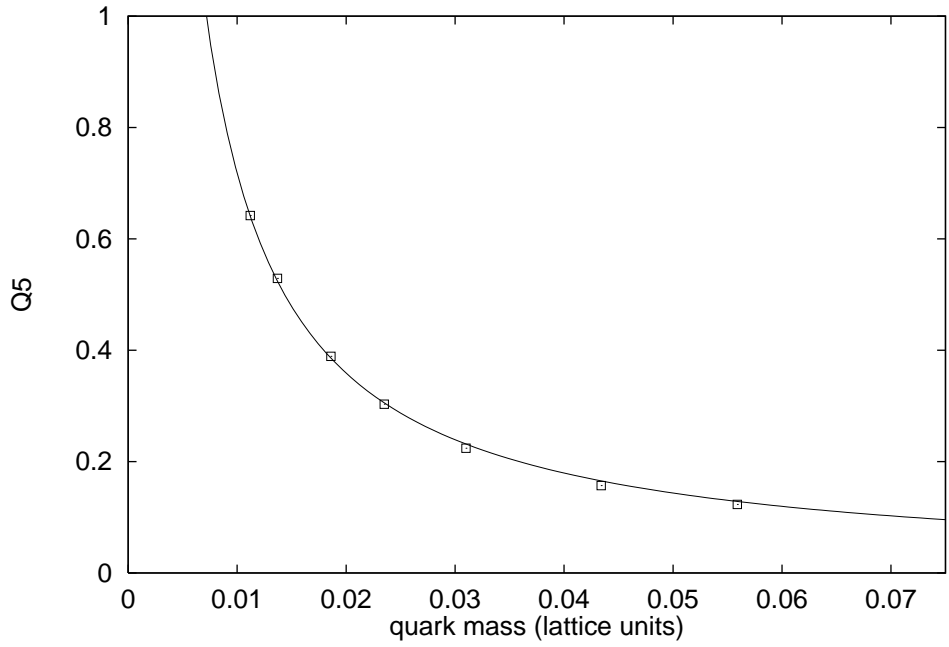}
\vspace{9.5cm}
\includegraphics{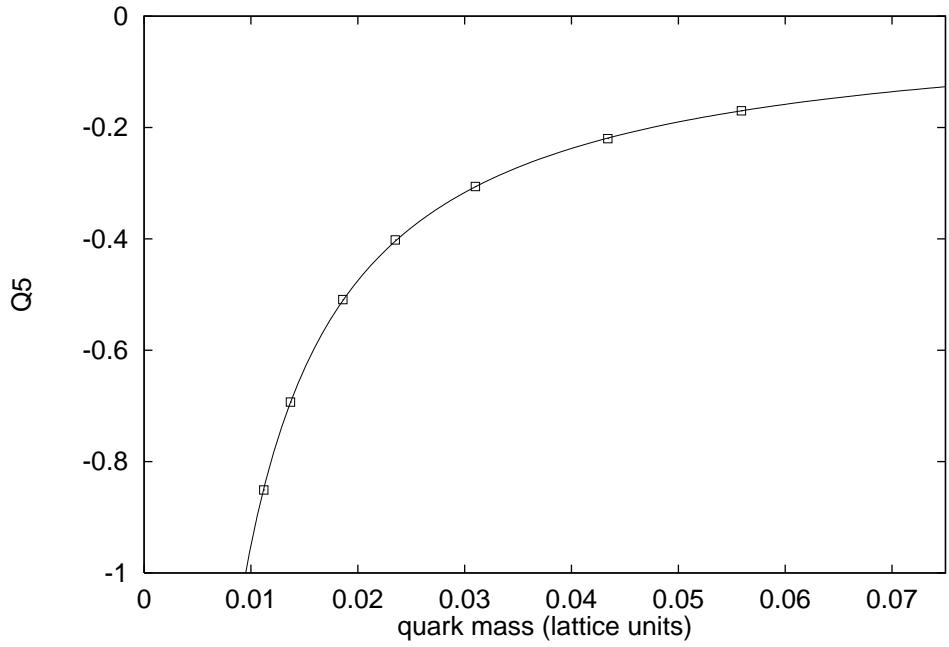}
\vspace{6.5cm}
\caption[]{$Q_5$ as a function of quark mass for two typical gauge configurations
after MQA pole shifting.}
\label{fig:q5mqa}
\end{figure}

\begin{figure}
\vspace*{2.0cm}
\includegraphics{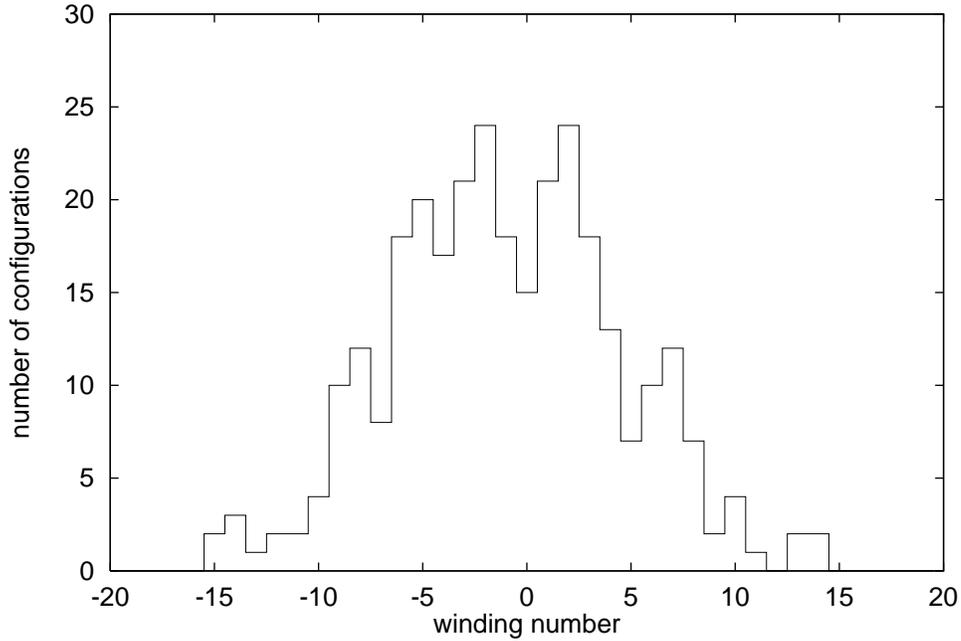}
\vspace{6.5cm}
\caption[]{The distribution of winding numbers determined from the integrated pseudoscalar 
density.}
\label{fig:windno}
\end{figure}

For both the hairpin correlator and the calculation of the integrated
pseudoscalar charge $Q_5$, the method used is one introduced into such
loop calculations in Ref.\cite{Kuramashi}. This method employs an ``allsource'' 
quark propagator calculated with a source that consists of a color-spin unit vector on
{\it all sites} of the lattice. This allows closed quark loops originating from
any space-time point to be included in a calculation (e.g. of $Q_5=Tr G\gamma^5$ or
of a hairpin diagram), relying on random gauge phases to cancel out the
gauge-variant open loops. Even on a single gauge configuration, this 
method is a reasonably accurate way of calculating global quantities like
$Q_5$, since random phase cancellation of noninvariant terms should take 
place in the sum over sites. 
The topological winding number $\nu$ of each gauge configuration can be determined
using the integrated anomaly equation \cite{SmitVink},
\begin{equation}
\nu = -im_qQ_5
\end{equation}
Thus we expect the quark-mass dependence of $Q_5$ to exhibit a simple pole at
$m_q = 0$ with residue given by the winding number. The plots of $Q_5$ vs.
$m_q$ for two typical configurations (after MQA pole shifting) are shown in Fig[\ref{fig:q5mqa}]. The solid
line in each case is the best single-pole fit. For the 300 lattice ensemble
with clover action, the distribution of winding numbers determined in this way
is shown in Fig[\ref{fig:windno}].

\begin{figure}
\vspace*{2.0cm}
\includegraphics{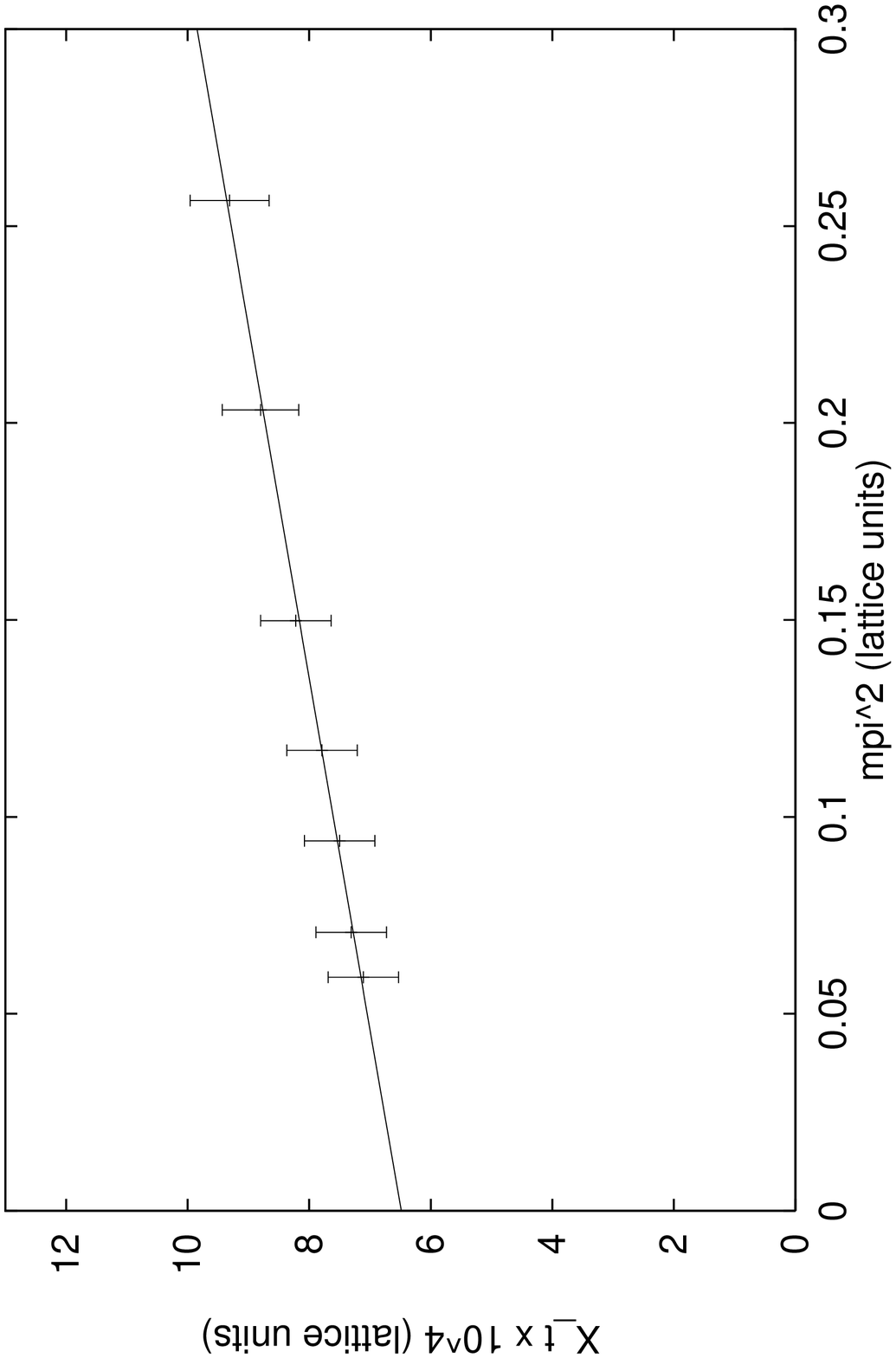}
\vspace{6.5cm}
\caption[]{Topological susceptibility calculated by the fermionic method on a $12^3\times 24$
lattice with clover improved ($C_{sw}=1.57$) fermions 
as a function of the quark mass used to determine $Q_5$.}
\label{fig:topsusc}
\end{figure}

The topological susceptibility $\chi_t$ can now be calculated
as the mean squared winding number per unit four-volume
\begin{equation}
\label{eq:topsusc}
\chi_t = \frac{\langle \nu^2\rangle}{V} = \frac{m_q^2}{V}\langle Q_5^2\rangle
\end{equation}
By calculating the last expression in (\ref{eq:topsusc}) we observe only a
slight quark-mass dependence of the result, as shown in Fig[\ref{fig:topsusc}]. Extrapolating
to the chiral limit, we obtain, for $C_{sw}=1.57$, 
\begin{equation}
\chi_t = 6.48(58)\times 10^{-4} = (188\; {\rm MeV})^4
\end{equation}
where the first result is in lattice units, and the second is obtained by using the
charmonium scale at $\beta=5.7$ of $a^{-1}=1.18$ GeV.
The corresponding result for unimproved Wilson fermions ($C_{sw}=0$) is
\begin{equation}
\chi_t = 3.24(39)\times 10^{-4} = (158\; {\rm MeV})^4
\end{equation}
Using the topological susceptibility and the value of axial vector decay constant
obtained from the valence propagator fits (see Section 6), the Witten-Veneziano
formula gives
\begin{equation}
\label{eq:chit_clover}
\delta = .063(6)
\end{equation}
for clover improved quarks ($C_{sw}=1.57$), and
\begin{equation}
\label{eq:chit_wilson}
\delta = .074(9)
\end{equation}
for unimproved Wilson quarks. In (\ref{eq:chit_clover}) and (\ref{eq:chit_wilson}) we used
the bare quark mass obtained from the hopping parameter to determine winding numbers from
$Q_5$ values. Here and elsewhere, we have taken the bare quark mass to be the 
pole mass 
\begin{equation}
 m_q = \log\left(1+\frac{1}{2}(\kappa^{-1}-\kappa_c^{-1})\right)
\end{equation}
but the results for the chiral log parameter are not significantly 
different if we use the naive bare mass $(\kappa^{-1}-\kappa_c^{-1})/2$.
If instead we use the current algebra mass 
\begin{equation}
m_q^{CA}\equiv f_Am_{\pi}^2/2f_P\;,
\end{equation}
the result (\ref{eq:chit_clover}) is essentially unchanged, while the 
$C_{sw}=0$ result (\ref{eq:chit_wilson}) is decreased to
\begin{equation}
\label{eq:chit_wilson2}
\delta = .032(4)
\end{equation}
(See discussion following Eq. (\ref{eq:hp_wilson}).)

\begin{figure}
\vspace*{2.0cm}
\includegraphics{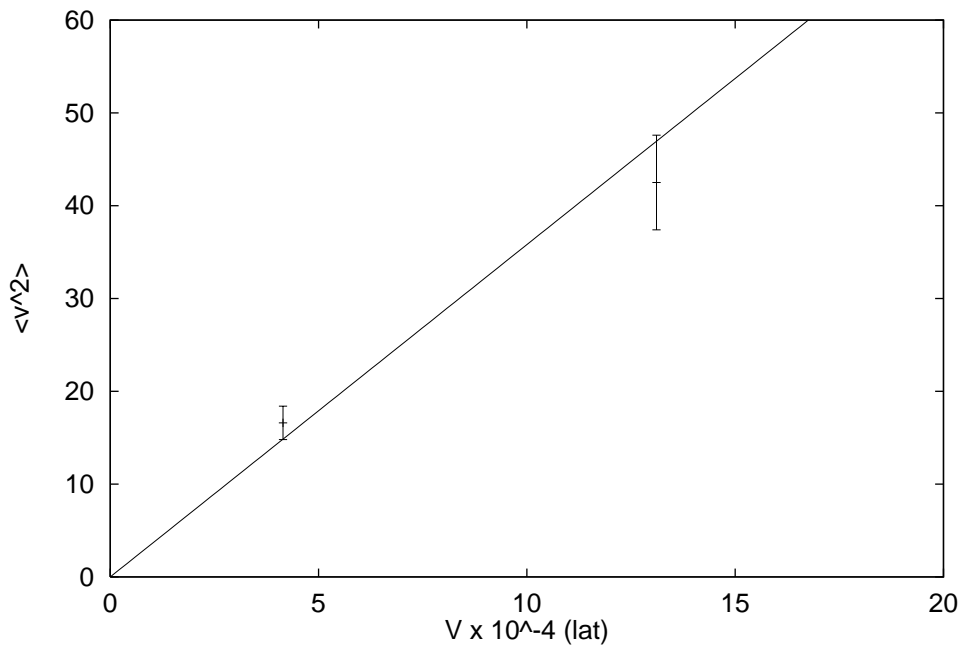}
\vspace{6.5cm}
\caption[]{The volume dependence of $\langle\nu^2\rangle$ for $12^3\times 24$ and
$16^3\times 32$ lattices, for Wilson-Dirac fermions, $C_{sw}=0$}
\label{fig:voldep}
\end{figure}

An important check on the calculation of topological susceptibility is to 
show that $\langle\nu^2\rangle$ has the proper dependence on volume,
i.e. $\langle\nu^2\rangle \propto V$. In Fig[\ref{fig:voldep}] we compare the topological
susceptibility calculated on a $12^3\times 24$ lattice and on a $16^3\times 32$
lattice, both with unimproved Wilson fermions ($C_{sw}=0$). 
The results shown in Fig[\ref{fig:voldep}] are consistent with the expected extensive property,
 i.e. a linear volume dependence.

\subsection{Hairpin correlator and the $\eta'$ mass insertion}

\begin{figure}
\vspace*{2.0cm}
\includegraphics{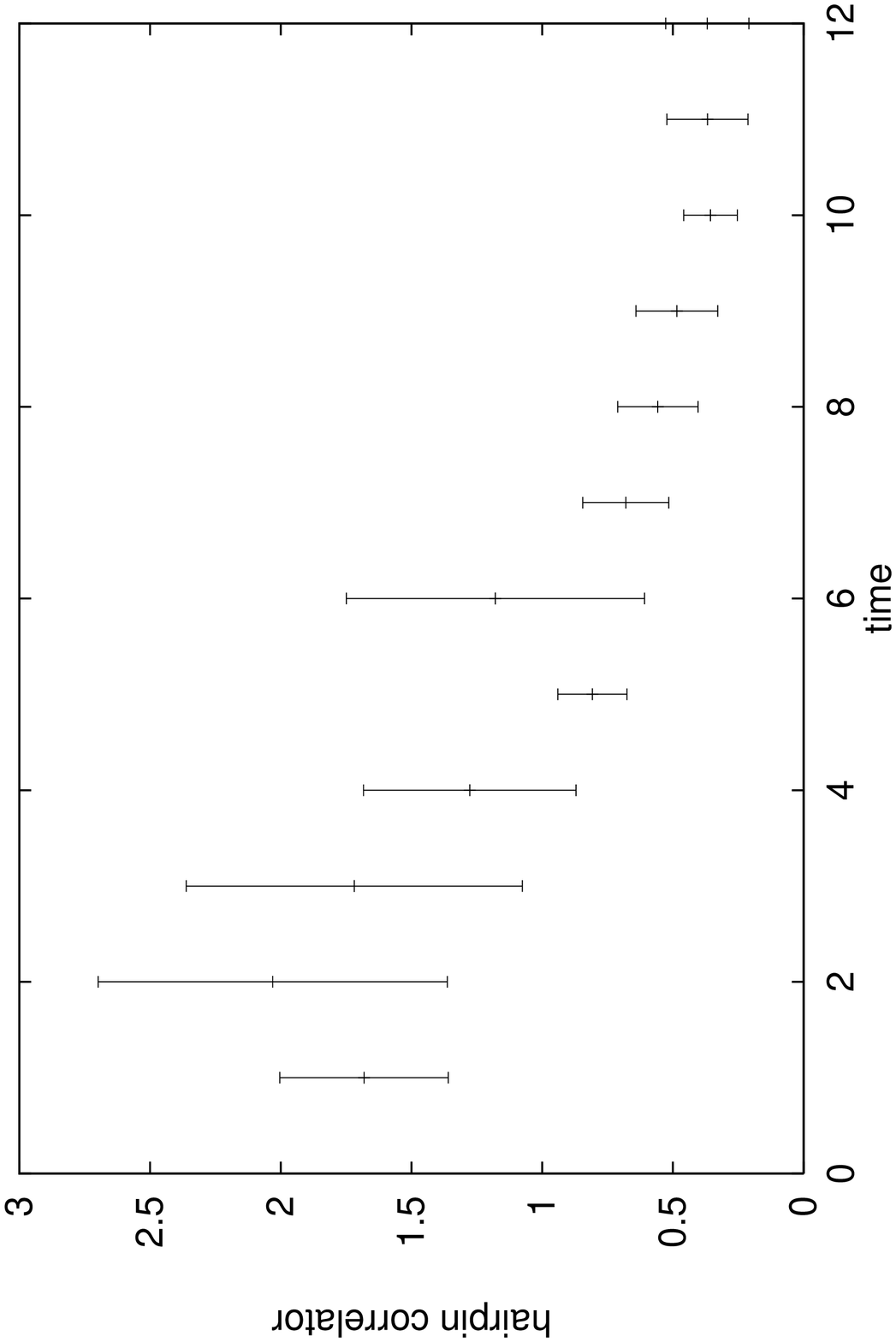}
\vspace*{9.5cm}
\includegraphics{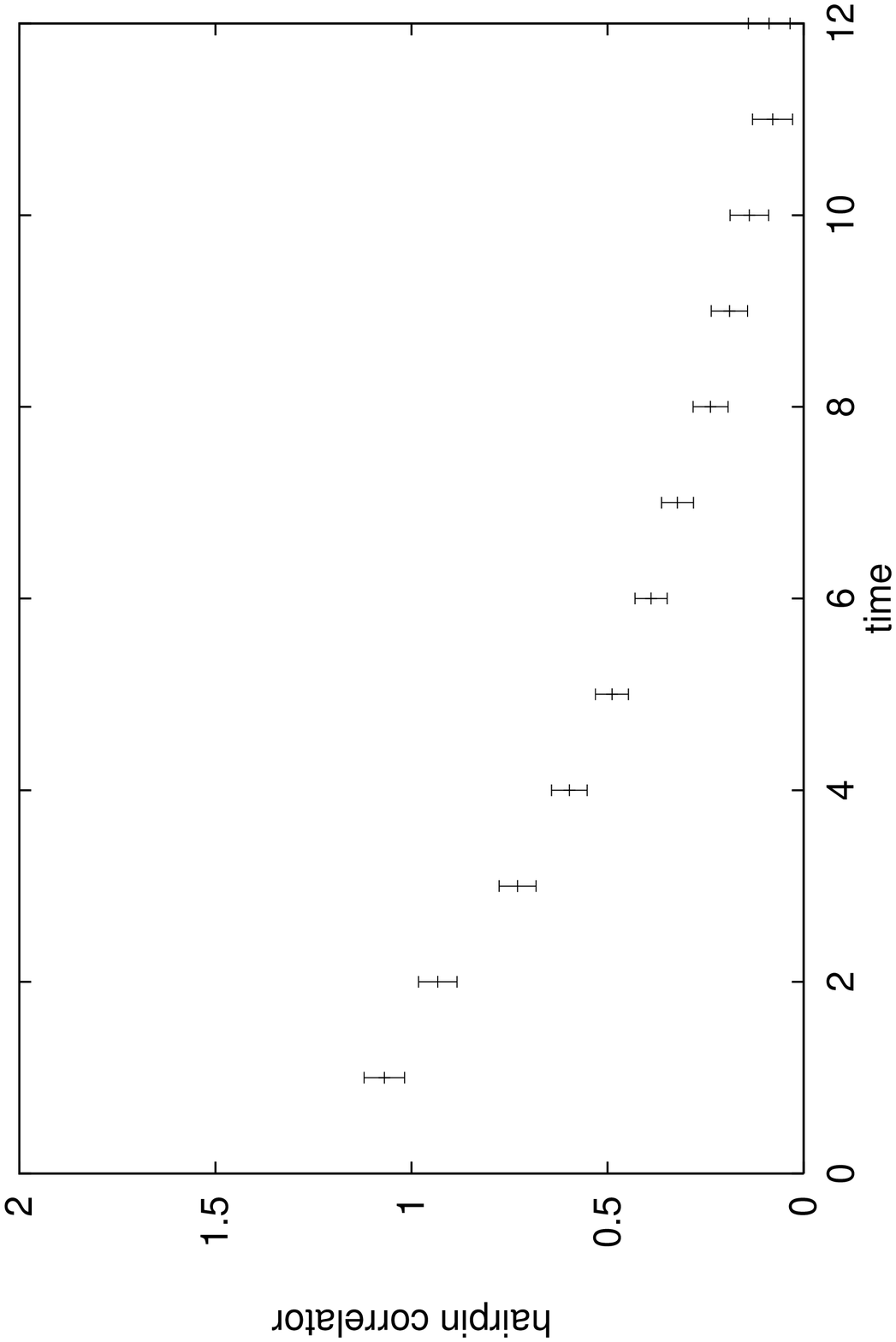}
\vspace{6.5cm}
\caption[]{The hairpin correlator for $\kappa=.1420$, $C_{sw}=1.57$ before and after
MQA improvement.}
\label{fig:haircorr}
\end{figure}

Using the same allsource propagators as in the previous subsection, we calculate
the hairpin contribution to the flavor singlet pseudoscalar propagator, i.e.
the loop-loop correlator (\ref{eq:hairpin}). Earlier calculations of this
correlator \cite{Kuramashi,lat96,Kilcup} were restricted to relatively heavy
quark mass and had large errors which prevented a detailed study of time-dependence.
As discussed in the Introduction,
the statistical problems encountered in these previous investigations
arise from exactly real Wilson-Dirac eigenmodes, the effect of which is magnified by the fact
that the hairpin propagator receives its largest contributions from topologically nontrivial gauge
configurations which necessarily contain such real modes. 
The MQA pole-shifting procedure is thus particularly effective
in improving the hairpin calculation. Fig[\ref{fig:haircorr}] shows an example of a hairpin 
propagator before and
after MQA improvement of the corresponding quark propagators. The quark mass
is still rather heavy here ($m_q\simeq$ 36 MeV, $m_{\pi}=.386 a^{-1}\approx 450$ MeV). For even lighter quarks, the unimproved
hairpin propagator is unmeasurable, with errors $>100\%$, while the MQA
improved hairpin is still quite well determined, allowing a reasonably accurate measurement
of the $\eta'$ mass insertion even at the lightest quark mass we have studied.

The size and time-dependence of the hairpin correlator is measured accurately enough
in the MQA method to address the two issues of excited state contamination and 
$p^2$-dependent vertex insertion terms mentioned in the Introduction. These 
two effects are distinct, but they are difficult to disentangle
from the time-dependence alone, since they both have the effect of adding
a single pole term to the correlator. Fortunately,
there is another way to determine the presence or absence of excited states,
namely, to study the ratio of hairpin correlators obtained from smeared and local
$\bar{\psi}\gamma^5 \psi$ sources. This can be compared with the overlap of the same smeared
and local sources with the ground-state pion, as determined from the large-$t$ behavior
of the corresponding valence pion propagators.

We construct smeared-source hairpin correlators by a modification
of the  allsource method used for the local-source hairpins. In the latter,
the source used for propagator inversion was a unit color-spin vector on every site.
In order to obtain meaningful results for smeared source propagators, we must
perform the smearing in Coulomb gauge.
The smeared sources used for the valence pion propagators were constructed 
using an exponential smearing function $\propto e^{-\lambda r}$.
Based on other studies of hadron wave functions at this value of $\beta$, we took
$\lambda = 0.5$ in lattice units. There is an additional subtlety in the implementation
of Coulomb gauge smearing in the allsource method. Since this method relies on random
gauge phase cancellations, the actual sums over sites for the two ends of the hairpin
must be carried out in the original unfixed gauge. In fact we carry out all calculations
in the unfixed gauge, just as in the local calculation. The only difference
is that the source used for propagator inversion is a ``smeared allsource'' which is
constructed by the following procedure:
\begin{enumerate}
\item  Construct an ordinary allsource, i.e. a unit
color-spin vector on every site.
\item To the allsource, apply the gauge transformation
that transforms from the original unfixed gauge to the Coulomb gauge.
\item  Smear the
source terms on each site in Coulomb gauge by convoluting with an exponential smearing function.
(This is most efficiently done in momentum space using FFT's.) 
\item  Transform the
smeared allsource back to the original unfixed gauge. 
\end{enumerate}

This smeared allsource can 
be used as the source for the quark propagator calculation, and the subsequent 
analysis is identical to that of the local hairpin correlator. In Coulomb gauge
the smeared allsource is a superposition of real exponential sources originating 
from every point. By going back to the unfixed gauge, we attach a random SU(3) gauge
phase to each exponential, so that a quark loop which starts on one exponential
and ends on another will have a random phase (even if it actually starts and ends at
the same space-time point), whereas terms which start and end on the same exponential have no
random phase (even if they start and end on different points). In this sense, the
method is very similar in spirit to one introduced earlier 
by Weingarten, et al \cite{Wein}, where 
multiple smeared sources are introduced in hadron spectroscopy calculations by attaching
random U(1) phases to each source.

The ratio of ground-state overlaps of the smeared and local sources with the pion 
is easily and accurately determined from the behavior of the smeared-smeared
and smeared-local valence pion propagators. As discussed in the next section,
values for the pion mass and for the ground-state overlaps are obtained from
a combined fit to the propagators using smeared pseudoscalar, local pseudoscalar,
and local axial vector sources and sinks.

\begin{figure}
\vspace*{2.0cm}
\includegraphics{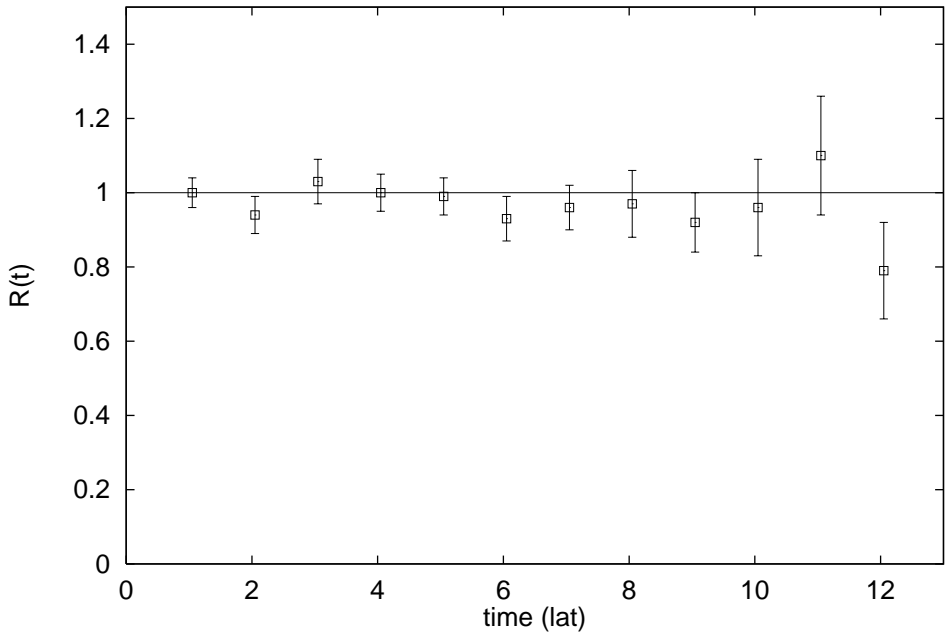}
\vspace{6.5cm}
\caption[]{Ratio $R(t)$ of the local-source hairpin divided by the smeared-source
hairpin with $\kappa=.1427$, normalized by the asymptotic valence propagator ratio.}
\label{fig:hairratio1427}
\end{figure}

\begin{figure}
\vspace*{2.0cm}
\includegraphics{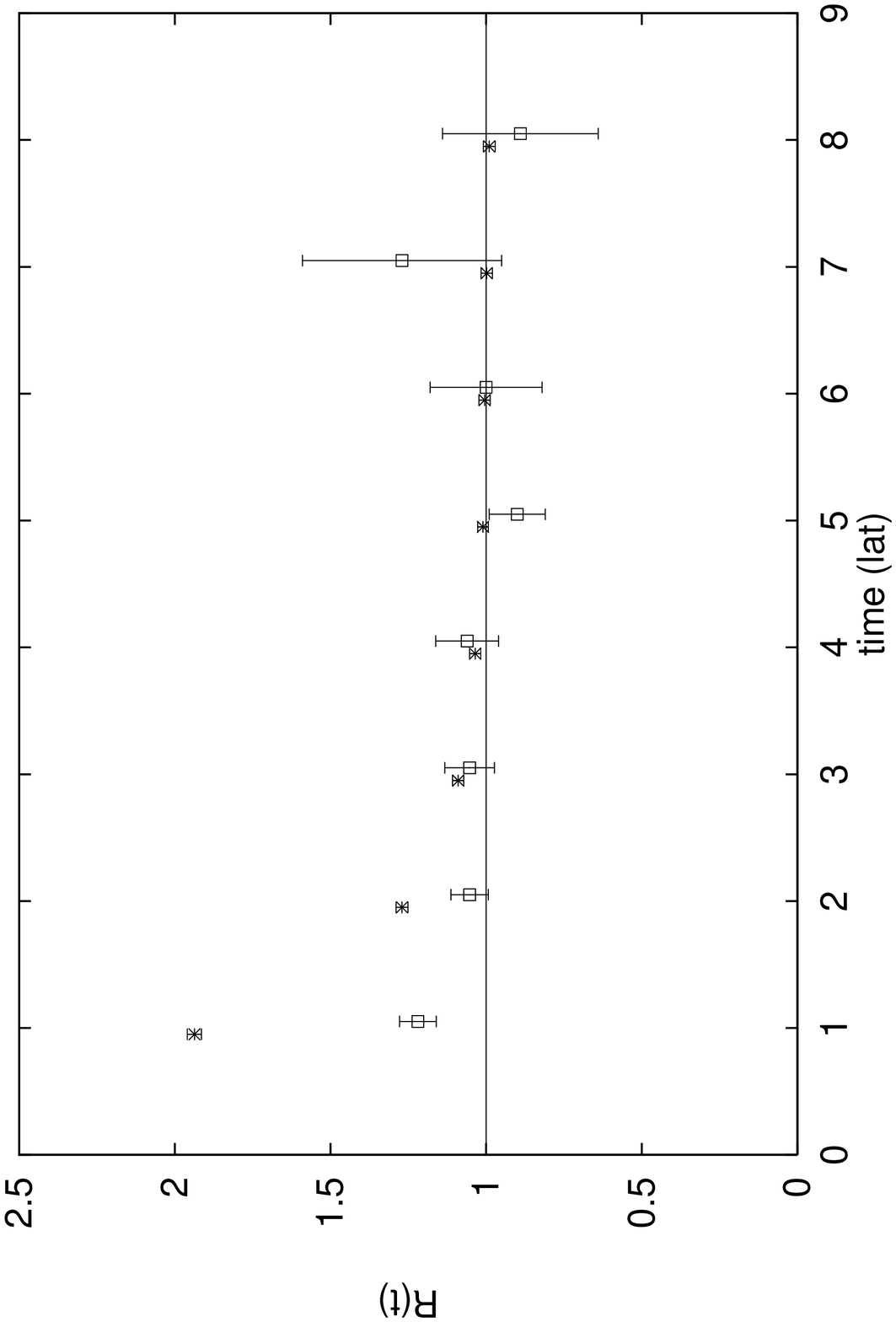}
\vspace{6.5cm}
\caption[]{Ratio $R(t)$ (boxes) of the local-source hairpin divided by the smeared-source
hairpin with $\kappa=.1410$, normalized by the asymptotic valence propagator ratio.
Also shown ($\times$'s) is the same ratio for the valence propagator.  }
\label{fig:hairratio1410}
\end{figure}

  Define the local and smeared $\bar{\psi}\gamma^5\psi$
operators by $P_l(x)$ and $P_s(x)$ respectively, and measure the corresponding
matrix elements,
\begin{equation}
\langle 0|P_i(0)|\pi(\vec{p}=0)\rangle = f_P^{(i)} \;\;\;\; i = l,s
\end{equation}
To test for the presence of excited states in the hairpin correlator, define
the smeared and local hairpin correlators (at zero 3-momentum) $\Delta_h^{(i)}(t),
\; i=s,l$ and
plot the ratio
\begin{equation}
R(t) \equiv \frac{\Delta_h^{(l)}(t)/(f_P^{(l)})^2}{\Delta_h^{(s)}(t)/(f_P^{(s)})^2}
\end{equation}
If there are no excited states, this ratio should be equal to unity.
In Fig[\ref{fig:hairratio1427}]
we plot this ratio for one of the lightest mass hairpins calculated ($\kappa=.1427$
or $m_q=.0137$ in lattice units).
The absence of any excited state contamination in the hairpin propagator is striking.
By contrast, the ratio of valence propagators at small times is substantially larger 
than its asymptotic value, indicating that the local valence propagator has a larger excited
state contribution. We conclude that the hairpin vertex
is very nearly decoupled from excited pseudoscalar states. In Fig[\ref{fig:hairratio1410}], we also 
plot the results of a similar analysis for a heavier quark mass ($k=.1410$ or
$m_q=.0559$ in lattice units). Here the relative contribution of excited states
to the valence propagator (also shown in the plot) is even larger than in the light mass case.
The hairpin propagator, on the other hand, still exhibits little if any excited
state contribution. For $t\geq 2$, there is no significant departure of the
hairpin ratio from its asymptotic value. This analysis of the smeared-to-local
hairpin ratio has been carried out at all the other mass values with similar
conclusions. In no case is there any significant indication of excited states
for $t\geq 2$.

The demonstrated absence of excited states from the hairpin diagram allows us to
make effective use of the time-dependence of the correlator to investigate
the structure of the hairpin vertex. The simplest assumption, often invoked
in phenomenological discussions, is that the hairpin is simply a momentum
independent mass insertion $m_0^2$. With this assumption, the quenched hairpin correlator
in momentum space is given by
\begin{equation}
\tilde{\Delta}_h(p) = f_P\frac{1}{p^2+m_{\pi}^2}m_0^2 \frac{1}{p^2+m_{\pi}^2}f_P
\end{equation}
Fourier transforming over $p_0$, this implies a time-dependence for the zero
momentum propagator of
\begin{equation}
\label{eq:doublepole}
\Delta_h(\vec{p}=0;t) = \frac{f_P^2m_0^2}{4m_{\pi}^3}(1+m_{\pi}t)e^{-m_{\pi}t} + (t\rightarrow N_{T}-t)
\end{equation}
This structureless hairpin vertex is suggested by large $N_c$ arguments, but
it is important to test for the more general possibility that the vertex has
some additional $p^2$ dependence. To lowest order in a $p^2$ expansion, this would
generalize the above analysis by the replacement
\begin{equation}
\label{eq:spdp}
m_0^2\rightarrow \Pi(p^2)\approx \Pi(-m_{\pi}^2)+(p^2+m_{\pi}^2)\Pi'(-m_{\pi}^2)+\ldots
\end{equation}

\begin{figure}
\vspace*{2.0cm}
\includegraphics{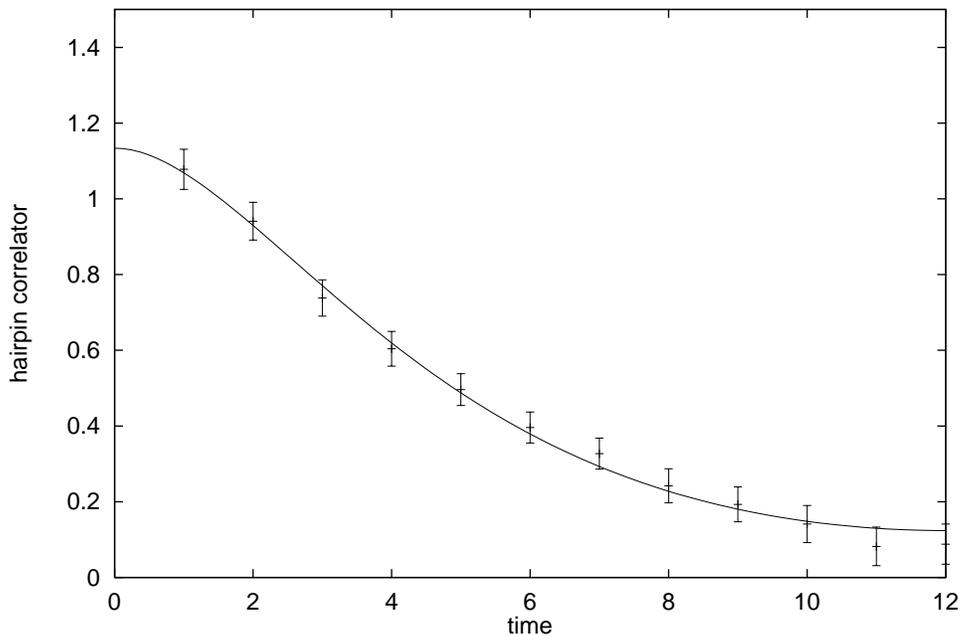}
\vspace{6.5cm}
\caption[]{One-parameter fit to a pure double-Goldstone pole form for the hairpin correlator at $\kappa=.1420$. Pion mass is fixed from the valence propagator analysis.  }
\label{fig:doublefit}
\end{figure}

To test for $p^2$ dependence of the hairpin vertex, and to estimate its effect on
the determination of the $\eta'$ mass, we carried out two sets of correlated fits
to the hairpin time-dependence, one with the pure double-pole formula (\ref{eq:doublepole})
and one to the single-pole + double-pole formula resulting from (\ref{eq:spdp}). In all these fits,
the pion mass was held fixed at the value given by the valence propagator analysis.
Since we have already demonstrated that there is very little excited state contamination
in the hairpin propagator, the range of times used in the fits is taken to be 
$2\leq t\leq 10$.  
To summarize the overall results of these fits, the hairpin time dependence for all
the quark masses studied is well described by a single-parameter fit to the pure 
double-pole formula (\ref{eq:doublepole}). (Here the pion mass is not a fit parameter,
since it is already accurately determined from the valence propagator.)
In Fig[\ref{fig:doublefit}] we show an example of a  pure double pole fit to the
hairpin correlator over the entire accessible t range.
We conclude that the hairpin vertex is reasonably
well described by a momentum-independent mass insertion.
The final results for $m_0^2$, given in Table 1, are extracted from pure double
pole fits. Using Equation (\ref{eq:delta}) and the lattice value for $f_{\pi}(=f_A/\sqrt{2})$
(see Table 2),
this gives
\begin{equation}
\label{eq:hp_clover}
\delta = .062(7)
\end{equation}
for $C_{sw}=1.57$, and
\begin{equation}
\label{eq:hp_wilson}
\delta = .044(5)
\end{equation}
for $C_{sw}=0$. 
Note that, in the pure double-pole approximation to the 
hairpin correlator, the value of $\delta$ from $\chi_t$ and that obtained from the
hairpin residue are related by a factor $(f_Am_{\pi}^2/2f_Pm_q)^2$, which should be
unity by the chiral Ward identity. The agreement between (\ref{eq:chit_clover}) and
(\ref{eq:hp_clover}) for the clover-improved calculations can be traced to the following two
facts: (1) The double-pole formula gives a good description of the hairpin correlator for all
time separations, and (2) The Ward identity is well-satisfied for $C_{sw}=1.57$. (By contrast,
the current algebra quark mass is about $40\%$ smaller than the bare mass for $C_{sw}=0$.)

\begin{table}
\centering
\caption{Value of $m_0$, the hairpin contribution to the $\eta'$ mass, 
for $C_{sw}=1.57$ (first two columns) 
and $C_{sw}=0$ (last two columns) at $\beta=5.7$, for $N_f$=3.}
\label{tab:hairpin}
\begin{tabular}{|c|c||c|c|}
\hline 
$\kappa$ & $m_0a$ & $\kappa$ & $m_0a$  \\
\hline
.1410 & .517(23) &       &          \\
.1415 & .534(24) & .1630 & .179(10) \\
.1420 & .554(24) & .1650 & .239(16) \\
.1423 & .568(26) & .1667 & .289(13) \\
.1425 & .576(27) & .1675 & .321(13) \\
.1427 & .576(30) & .1680 & .338(13) \\
.1428 & .576(33) & .1685 & .353(18)\\
$\kappa_c$ & .580(27) & $\kappa_c$ & .393(15) \\
\hline
\end{tabular}
\end{table}

\newpage
\section{Quenched chiral logs in the pseudoscalar mass}

The effect of quenched $\eta'$ loops on the chiral behavior of the pseudoscalar 
mass is one of the most definitive predictions of the quenched chiral log 
analysis.\cite{Sharpe,B&G} In a previous analysis of the pion mass as a function 
of quark mass,\cite{lat96} no evidence was found for quenched chiral log behavior 
at $\beta=5.7$ for unimproved Wilson fermions, 
with a one-standard-deviation upper bound on the chiral log parameter 
 of $\delta<.07$. This was also shown to be consistent with the size of the hairpin 
propagator. That analysis was done before the development of the 
MQA method for resolving the exceptional 
configuration problem, and the lightest pion mass used was $m_{\pi}a=.253$ 
(hopping parameter $\kappa=.1680$ and $C_{sw}=0$).  With MQA improvement of quark 
propagators we obtain much better statistical errors on $m_{\pi}$ and also are able 
to go to a much lighter quark mass ($m_{\pi}a=.164$ or $\kappa=.1687$ for $C_{sw}=0$ 
and $m_{\pi}a=.244$ or $\kappa=.1428$ for $C_{sw}=1.57$). As we discuss in this section, 
this improved analysis allows us to observe clearly the quenched chiral log effect 
in the pion mass with a value of the chiral log parameter $\delta$ slightly less than 
the previously established upper bound ($\delta=.054$) for the $C_{sw}=0$ case. 
The value of $\delta$ is somewhat larger for clover improved quarks ($\delta=.073$), 
suggesting that the suppression of $\delta$ compared to the 
expected continuum value $\approx .17$ may be at least partially due to finite lattice spacing effects. 
The recent results from CPPACS of $\delta=.06-.12$\cite{CPPACS} for several values of $\beta$ 
is consistent with this possibility but is not accurate enough to observe any clear 
lattice spacing dependence.

To extract a value of the chiral log parameter from the pion mass as a function of 
bare quark mass, we shall consider in this section only the case where the two quark masses involved 
are equal (in Section 7  chiral log formulas for unequal quark masses will be derived from
a model chiral Lagrangian and   
 used to perform global fits to the b lattice data). 
In the equal quark mass case, the chiral logs sum up in a leading log approximation
to give an anomalous power law dependence of the squared pion mass on the quark mass, 
\begin{equation}
\label{eq:pimass}
m_{\pi}^2 = {\rm const.}\times m_q^{\frac{1}{1+\delta}}
\end{equation}
For $C_{sw}=1.57$ we calculated the pion mass at 9 values of hopping parameter 
ranging from $\kappa=.1400$ to .1428. The masses are obtained from a combined, 
correlated fit of smeared-local and smeared-smeared propagators, using a smeared 
pseudoscalar source, a local pseudoscalar source, and a local axial-vector source. 
The pion masses obtained are listed in Table 2. 

\begin{table}
\centering
\caption{Pseudoscalar masses and decay constants for $C_{sw}=1.57$ (first four columns) 
and $C_{sw}=0$ (last four columns) at $\beta=5.7$.}
\label{tab:masses}
\begin{tabular}{|c|c|c|c||c|c|c|c|}
\hline 
$\kappa$ & $m_Pa$ & $f_Aa$ & $f_Pa$ & $\kappa$ & $m_Pa$ & $f_Aa$ & $f_Pa$ \\
\hline
.1400 & .603(2) & .196(2) & .458(5)  & .1610 & .647(1) & .224(3) & .518(9) \\
.1405 & .556(2) & .190(2) & .444(5)  & .1630 & .558(1) & .199(2) & .471(9)  \\
.1410 & .505(2) & .183(2) & .430(6)  & .1650 & .458(1) & .174(2) & .424(8)  \\
.1415 & .450(3) & .176(2) & .418(6)  & .1667 & .356(2) & .154(2) & .387(8)  \\
.1420 & .386(3) & .169(2) & .410(7) & .1675 & .297(2) & .144(2) & .371(8)  \\
.1423 & .342(4) & .165(3) & .410(9) & .1680 & .254(2) & .136(5) & .358(12) \\
.1425 & .307(4) & .163(3) & .413(10) & .1683 & .221(3) & .132(6) & .353(19) \\
.1427 & .267(5) & .161(4) & .424(14) & .1685 & .195(4) & .129(7) & .342(25) \\
.1428 & .245(6) & .161(5) & .439(17) & .1687 & .164(5) & .126(11) & .345(38) \\
$\kappa_c$ & $---$ & .151(2) & $---$ &$\kappa_c$ & $---$ & .122(2) & $---$ \\
\hline
\end{tabular}
\end{table}

\begin{figure}
\vspace*{2.0cm}
\includegraphics{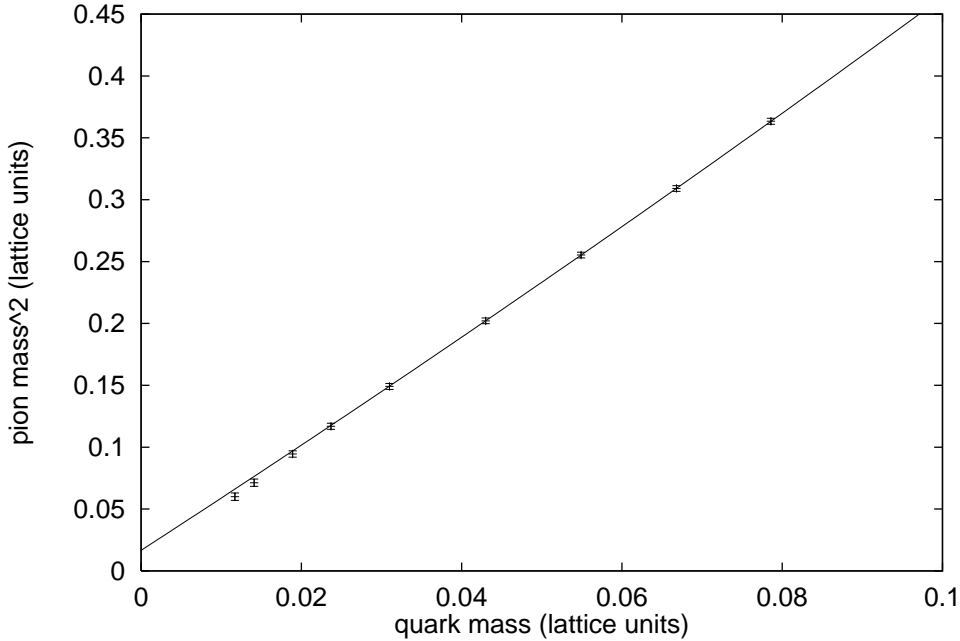}
\vspace{6.5cm}
\caption[]{$m_P^2$ vs. quark mass for $\beta=5.7$ and $C_{sw}=1.57$. Solid line is the second
order $\chi PT$ fit to the four heaviest masses.  }
\label{fig:mpivsmq}
\end{figure}

\begin{figure}
\vspace*{2.0cm}
\includegraphics{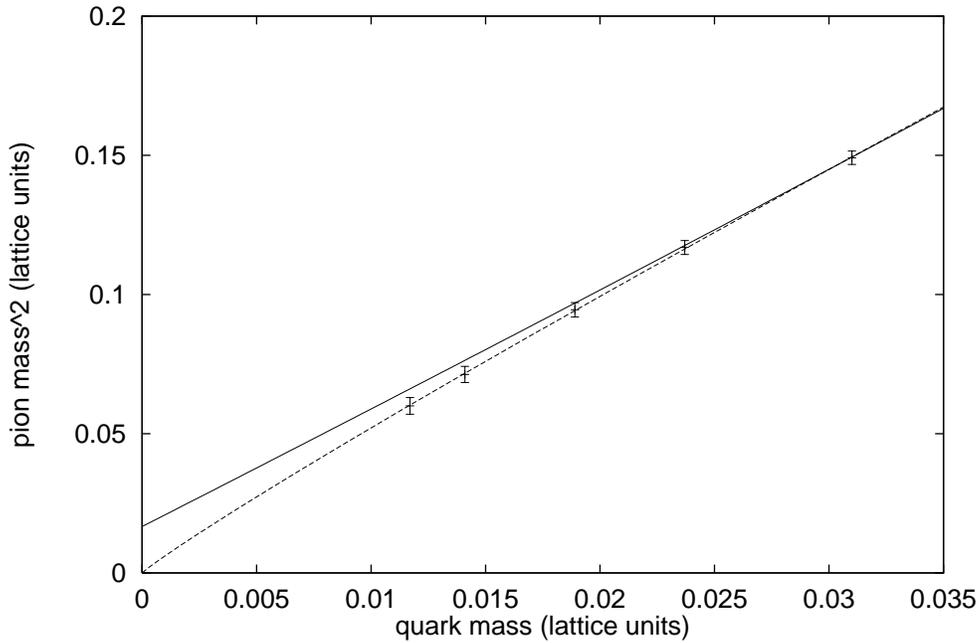}
\vspace{6.5cm}
\caption[]{Quenched chiral log fit (dashed line) to the lightest 5 mass values 
with $\delta = .073$. 
Solid line is the extrapolated $\chi PT$ fit to the heaviest four
masses.}
\label{fig:qnchlogfit}
\end{figure}

Figures [\ref{fig:mpivsmq}] and [\ref{fig:qnchlogfit}] exhibit the chiral log effect in the pseudoscalar mass 
graphically. The first plot includes 
all nine values of quark mass. The solid line is the best quadratic fit (i.e. $m_{\pi}^2=Am_q+Bm_q^2$)
to the four heaviest masses (with $\kappa_c$ included as a fit parameter).
The second plot  (Fig[\ref{fig:qnchlogfit}]) is an expanded view of the small mass region. It shows 
clearly that the light pion masses fall below the quadratic extrapolation of the 
heavier masses (solid line). The dashed line is a fit of the lowest five masses to the chiral 
log formula (\ref{eq:pimass}) (again with the value of $\kappa_c$ as one of the fit parameters). We find
\begin{equation}
\delta = .073(20)
\end{equation}

We have also calculated the pion masses for $C_{sw}=0$ on an ensemble of 200 gauge 
configurations at $\beta=5.7$ on a $16^3\times 32$ lattice (the ``a'' ensemble from 
the ACPMAPS library). Here also we calculate $m_{\pi}$ at nine values of hopping 
parameter ranging from $\kappa=.1610$ to $\kappa=.1687$. For this case we were 
able to go to an even smaller pion mass of $m_{\pi}a=.164$ which is less than
200 MeV in physical units (using the charmonium scale $a^{-1}=1.18$ GeV). 
The pion masses are listed in Table 2. 
Again the quenched chiral log
effect is clearly visible, with the lightest-mass points falling significantly below an 
extrapolated quadratic fit. Fitting to the quenched chiral log formula (\ref{eq:pimass})
we find
\begin{equation}
\delta = .054(20)
\end{equation}
Consistent with the direct hairpin calculation, the value of the chiral log parameter
from the $m_{\pi}^2$ analysis
is somewhat smaller for $C_{sw}=0$ than for $C_{sw}=1.57$, although in this case
the error bars are larger, so the difference is only marginally significant.

\newpage
\begin{figure}
\vspace*{2.0cm}
\includegraphics{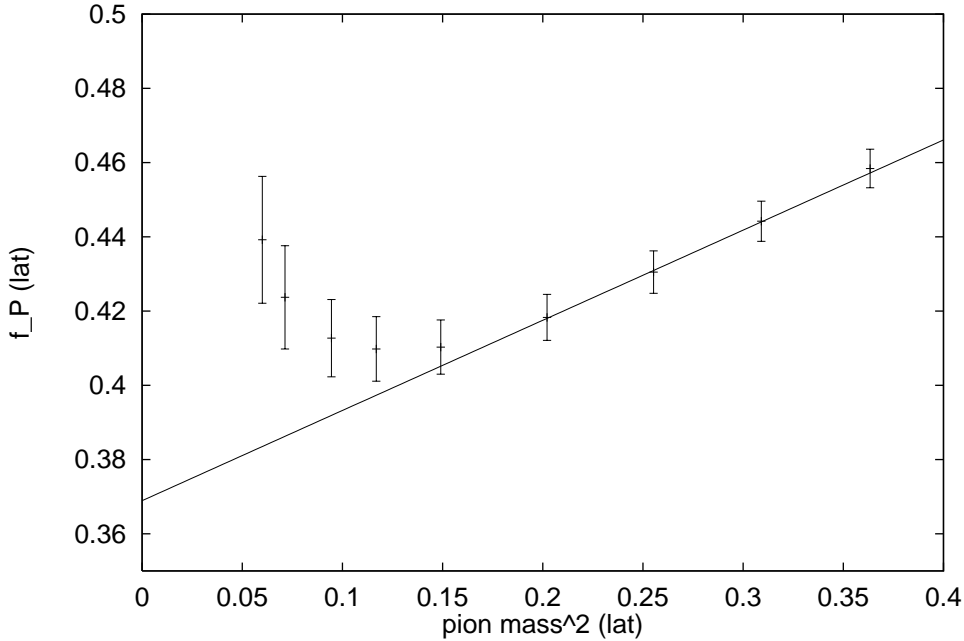}
\vspace{6.5cm}
\caption[]{ Pseudoscalar decay constant $f_P$ as a function of quark mass. Enhancement at
small $m_q$ is a QCL effect.  }
\label{fig:fpvsmq}
\end{figure}

\section{Chiral behavior of the pseudoscalar and axial-vector matrix elements}  

\begin{figure}
\vspace*{2.0cm}
\includegraphics{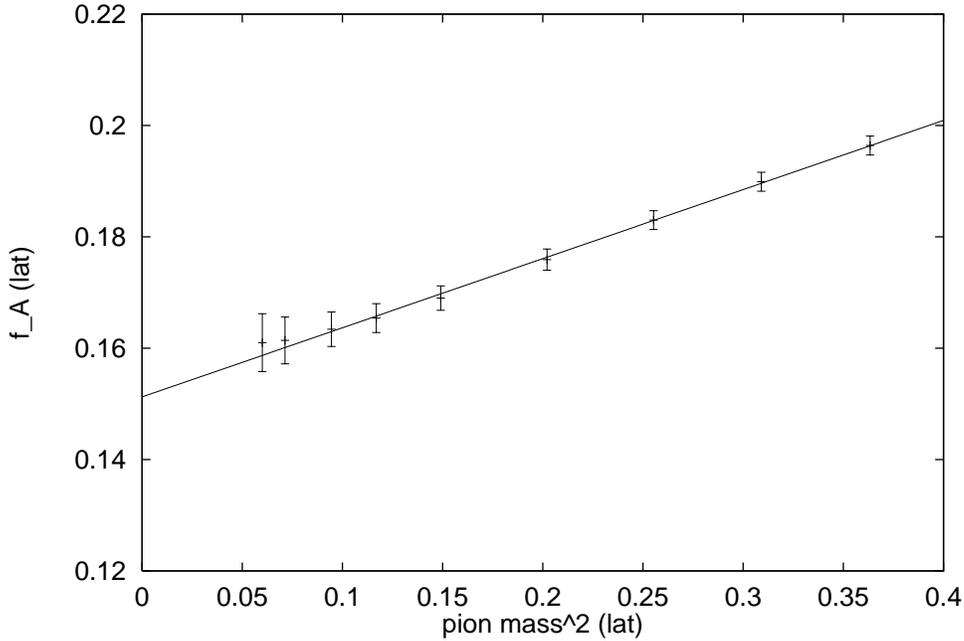}
\vspace{6.5cm}
\caption[]{Axial-vector decay constant $f_A$ as a function of quark mass. Note the 
absence of a QCL enhancement, as predicted by theoretical arguments.  }
\label{fig:favsmq}
\end{figure}

As discussed in the Introduction, the chiral behavior of the pseudoscalar and
axial-vector decay constants $f_P$ and $f_A$ provide further tests of quenched
chiral log predictions. When the two quarks in the pseudoscalar meson have equal mass, we
should find a clear contrast between these two quantities: $f_P$ should exhibit
a QCL factor $\propto (m_{\pi}^2)^{-\delta}$ while $f_A$ should have a smooth, nonsingular chiral
limit. The values of $f_P$ and $f_A$ are obtained from the combined fit to 
smeared-smeared and smeared-local propagators discussed in Section 5.
The effects of full tadpole improvement for the normalization of quark
masses and quark operators have been included in our calculations, but we have
 not carried out a complete nonperturbative O(a) improvement program.
The numerical results are presented in Table 2. For the $C_{sw}=1.57$ results, the
mass dependence of the decay constants is shown in Figures [\ref{fig:fpvsmq}] and
[\ref{fig:favsmq}]. It is clear
from these plots that at the very lightest masses, the value of $f_P$ is significantly
larger than a linear extrapolation of the heavier mass results, consistent with
the singular $(m_{\pi}^2)^{-\delta}$ expected from QCL effects, while
$f_A$ exhibits no sign of singular behavior, and is well described by a linear fit.
In both figures, the solid line is the best linear fit 
to the four heaviest masses.

\begin{figure}
\vspace*{2.0cm}
\includegraphics{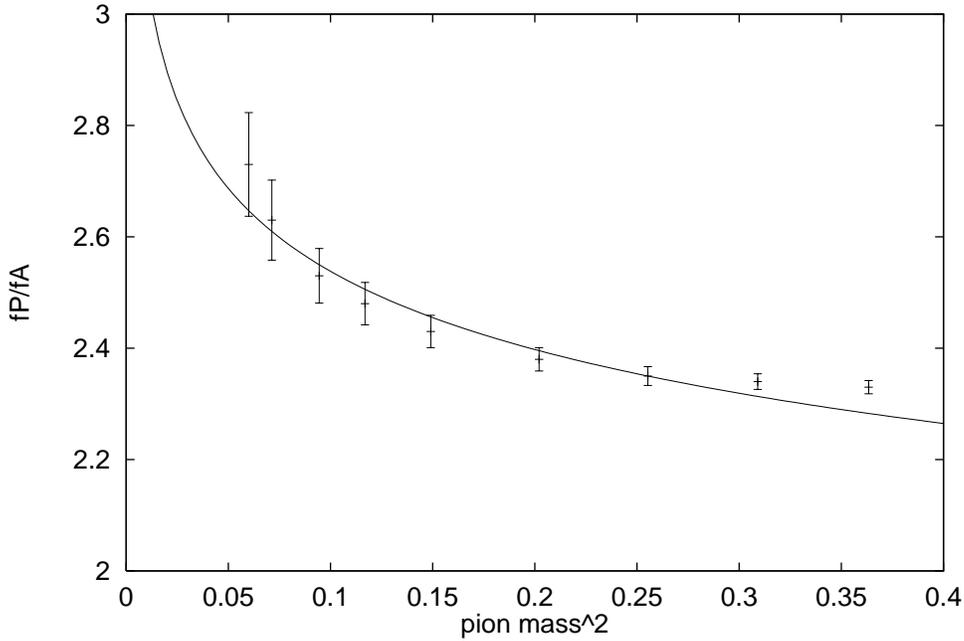}
\vspace{6.5cm}
\caption[]{ Ratio $f_P/f_A$ as a function of quark mass. Solid line is a QCL fit
with $\delta=.085$.  }
\label{fig:fpoverfa}
\end{figure}

Since the singular behavior of $f_P$ in the chiral limit is determined by the 
ubiquitous chiral log parameter $\delta$, it should be possible to use the results
for $f_P$ to obtain another estimate of this parameter. The extraction of a value
of $\delta$ is made somewhat more difficult by the fact that, in addition to
the singular QCL effect, the mass-dependence of $f_P$ also has a significant
contribution from higher order chiral perturbation theory (i.e. terms linear in
$m_{\pi}^2$). With the accuracy of
our data, the two contributions are difficult to disentangle. Although the QCL
effect is clearly visible, a fit to $f_P$
which includes both QCL terms and $\chi PT$ terms is rather unstable and the
resulting value for $\delta$ is poorly determined. We can do much better if
we make an additional, phenomenologically motivated assumption that the perturbative
slopes of $f_P$ and $f_A$ are approximately equal \cite{Leutwyler}. 
The data in Figures [\ref{fig:fpvsmq}] and [\ref{fig:favsmq}] are consistent with this assumption. The ratio
$f_P/f_A$ should thus exhibit a relatively pure chiral log behavior,
\begin{equation}
\label{eq:fpfa}
\frac{f_P}{f_A} ={\rm const.}\times (m_{\pi}^2)^{-\delta}
\end{equation}
The lattice results for this ratio are shown in Fig[\ref{fig:fpoverfa}], along with the best
fit to the QCL formula (\ref{eq:fpfa}). This gives a value of the chiral log
parameter of
\begin{equation}
\delta = .085(23)
\end{equation}
\newpage
\section{Extraction of $\delta$ from mass and decay constant cross-ratios}

 To facilitate a comparison with previous work by the CPPACS collaboration
 \cite{CPPACS}, we have extracted the chiral log parameter $\delta$ from
 our full set of b-lattice clover-improved results for masses and
 decay constants of the 45 independent mesons which can be formed from
 the nine available quark masses, using the cross-ratio method introduced
 in \cite{CPPACS}. For a given meson parameter $y_{ij}$ (here $i,j$ label
 the quarks in the meson and run from 1 to 9), the cross-ratio $R_{ij}$ is
 defined as follows
\begin{equation}
\label{eq:xratio}
   R_{ij} \equiv \frac{y_{ij}^{2}}{y_{ii}y_{jj}}
\end{equation}
 Let $M_{ij},f_{P;ij},$ and $f_{A;ij}$ denote the mass, pseudoscalar
 and axial-vector decay constant of the meson with quark content $i,j$
 (and quark masses $m_i$ and $m_j$).
 Then, with either $y_{ij}=M^{2}_{ij}/(m_i+m_j)$ or $y_{ij}=\frac{f_{P;ij}}{f_{A;ij}}$,
 one has, to leading order in $\delta$, {\em but ignoring higher order chiral
 perturbation theory contributions} (this amounts to setting $L_5,L_8=0$ in the
 chiral formulas (\ref{massformula},\ref{fpformula},\ref{faformula}) discussed
 in Section 8 below),
\begin{equation}
\label{eq:rij}
  R_{ij} = {\rm const.}\times (1 + \delta X_{ij})
\end{equation}
where the chiral logarithm is contained in the factor $X_{ij}$. At infinite
 volume, this factor becomes   
\begin{equation}
  X_{ij} = 2 + \frac{m_i+m_j}{m_i-m_j}\ln{\frac{m_j}{m_i}}
\end{equation}

\begin{figure}
\psfig{figure=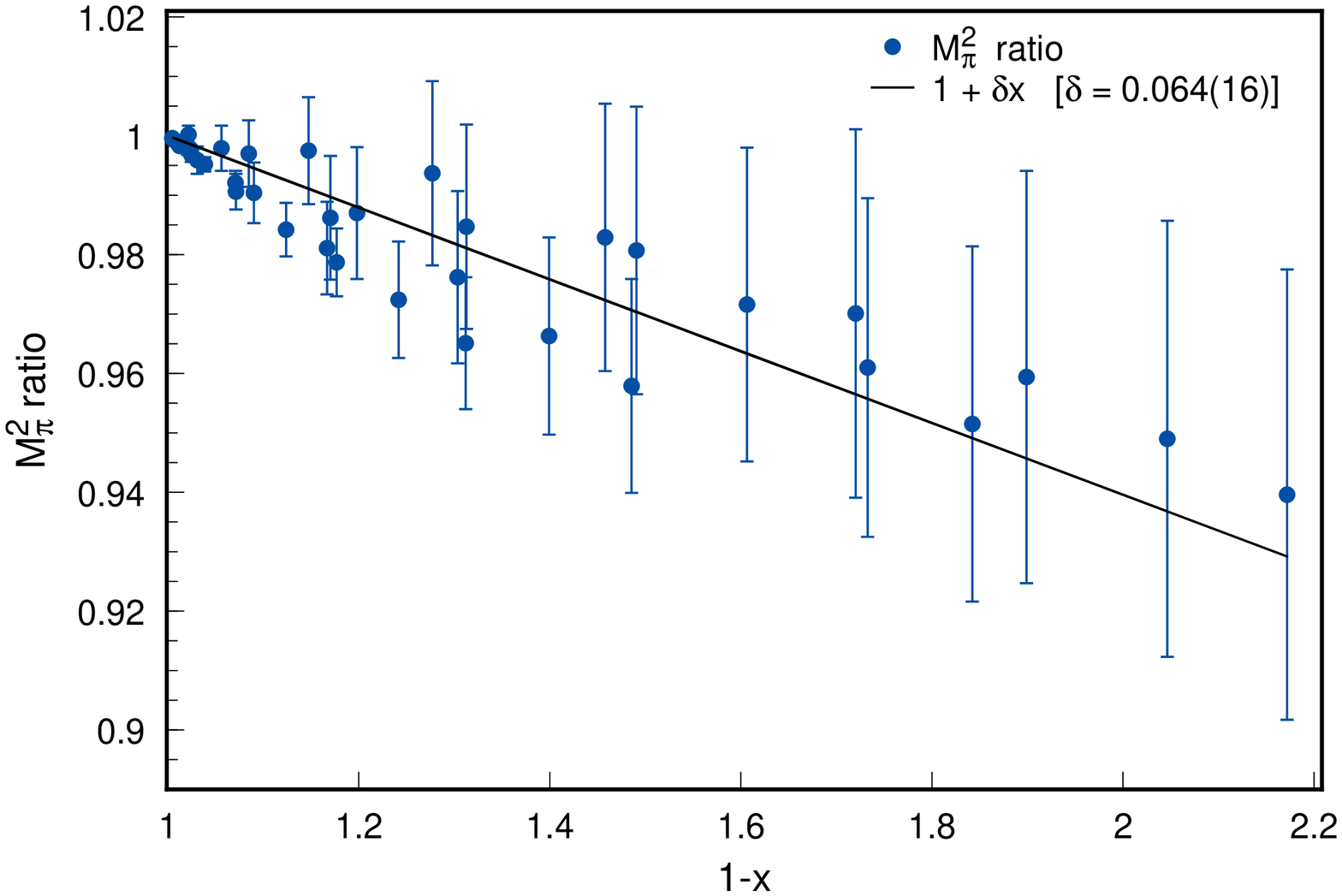,width=0.8\hsize}
\caption[]{ Cross-Ratios of  $M^{2}_{ij}/(m_i+m_j)$ as a function of finite volume $X_{ij}$ variable}
\label{fig:massratio}
\end{figure}

 Our lattices are at smaller volume that those in the work cited above
 (physical extent $\simeq$ 2 fermi as compared to 3 fermi in \cite{CPPACS}) and
 we go to considerably smaller quark masses, so we have used a 
 finite volume version of the fitting parameter $X_{ij}$ (see Section 8
 below for a discussion):
\begin{equation}
\label{xfinvol}
  X_{ij} = 2I_{ij}-I_{ii}-I_{jj}
\end{equation}
where the finite volume sums $I_{ij}$ are defined in (\ref{chintreg}).

\begin{figure}
\psfig{figure=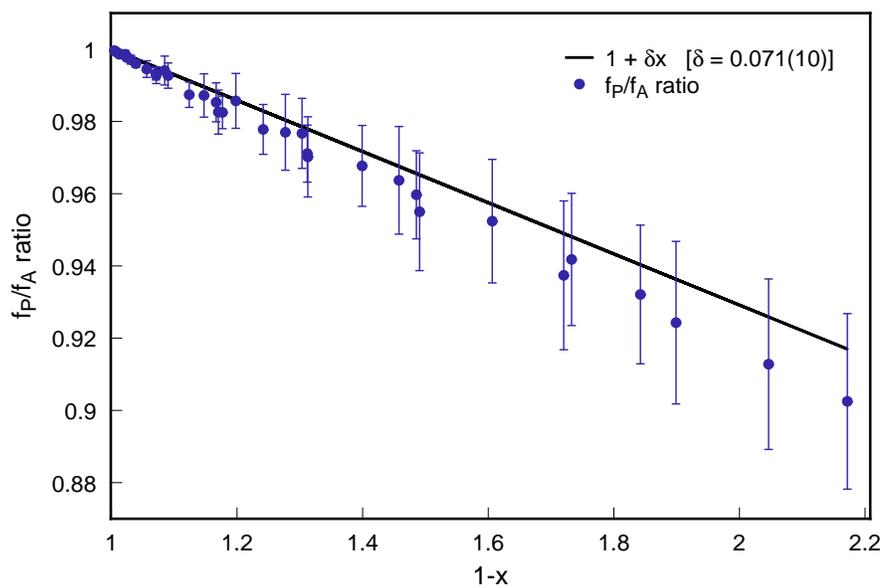,width=0.8\hsize}
\caption[]{ Cross-Ratios of  $f_{P;ij}/f_{A;ij}$ as a function of finite volume $X_{ij}$ variable}
\label{fig:fpfaratio}
\end{figure}

 Fitting the cross-ratios of the masses of  all 36 off-diagonal mesons, we obtain
\begin{equation}
  \delta = 0.0604 \pm 0.0155 \;\;\;(\chi^{2}=54/35\;{\rm d.o.f})
\end{equation}
 while a fit of cross-ratios of the decay constant ratios $\frac{f_{P;ij}}{f_{A;ij}}$ gives
\begin{equation}
 \delta = 0.0708 \pm 0.0104 \;\;\;(\chi^{2}=37/35\;{\rm d.o.f})
\end{equation}
 Finally, a combined fit of both the mass and decay-constant ratios gives
\begin{equation}
 \delta = 0.0732 \pm 0.0081 \;\;\;(\chi^{2}=118/71\;{\rm d.o.f})
\end{equation}

 The expected linearity in the X-variable of (\ref{eq:rij}) is displayed
 graphically in Fig[\ref{fig:massratio}] for the
 pseudoscalar masses and in Fig[\ref{fig:fpfaratio}] for the  ratio of pseudoscalar
 to axial decay constants. The results of fully correlated fits are displayed
 as solid lines.

\clearpage

\section{Comparison with quenched chiral perturbation theory}
For pions made from a quark and antiquark with unequal masses, the form of the
quenched chiral log effect is more complicated \cite{CPPACS,B&G}. For the range of masses we consider,
it is sufficient to keep only lowest order terms in $\delta$ (i.e. one-loop terms), or equivalently,
in a hairpin mass term which can be included explicitly as a correction to the basic
chiral Lagrangian :
\begin{eqnarray}
\label{eq:chilag}
 {\cal L}&=& \frac{f^2}{4}{\rm Tr}(\partial_{\mu}U^{\dagger}\partial^{\mu}U) \nonumber \\
 &+&\frac{f^2}{4}{\rm Tr}(
\chi^{\dagger}U+U^{\dagger}\chi)
   +L_5 {\rm Tr}(\partial_{\mu}U^{\dagger}\partial^{\mu}U(\chi^{\dagger}U+U^{\dagger}\chi)) \nonumber \\
 &+&L_8 {\rm Tr}(\chi^{\dagger}U\chi^{\dagger}U+U^{\dagger}\chi U^{\dagger}\chi) + {\cal L}_{\rm hairpin}
\end{eqnarray}
where
\begin{equation}  
 {\cal L}_{\rm hairpin} \equiv -\frac{1}{2} m_{0}^{2}\frac{f^2}{8}(i{\rm Tr}\ln(U^{\dagger})-i{\rm Tr}\ln(U))^2
\end{equation} 
The lowest order chiral Lagrangian has been supplemented by the
chiral symmetry breaking terms $L_{5}$ of
O($p^{2}m_{\pi}^{2}$) and $L_{8}$ of
  O($m_{\pi}^4$)  which model the leading mass-dependence of the
 slope in the pseudoscalar masses.
Starting from this Lagrangian, we can derive explicit formulas for the pseudoscalar masses,
and pseudoscalar and axial vector decay constants, consistent through order $p^4$ and including
the effects of the hairpin mass insertion (assumed local) through the term ${\cal L}_{\rm hairpin}$
in (\ref{eq:chilag}). The coefficients $L_{5}, L_{8}$ follow the notation of Gasser and Leutwyler \cite{Leutwyler}. 
The evaluation of the one loop chiral integrals appearing in this calculation
has also been carried out at finite volume (appropriate for the physical size of the lattices used),
and with a Pauli-Villars subtraction to regulate the ultraviolet. For example, logarithmically
divergent integrals such as

\begin{equation}
\label{chint}
  I_{ij} = \frac{1}{\pi^2}\int d^{4}p \frac{1}{p^2+M_{i}^2}\frac{1}{p^2+M_{j}^{2}}
\end{equation}
are replaced by
\begin{equation}
\label{chintreg}
  I_{ij} = 16\pi^{2}\sum_{p} (D(p,M_{i})D(p,M_{j})-D(p,\Lambda)^{2})
\end{equation}
where the momentum integration is now a discrete sum over a finite volume free boson
propagator $D(p,M)$. We have typically chosen the cutoff scale $\Lambda \simeq \frac{1}{a}$ but
 sensitivity of the results to this choice is very small. One also encounters quadratically
divergent graphs in the course of the calculation, which are regulated as follows:
\begin{equation}
 I_{i} = 16\pi^{2}\sum_{p} (D(p,M_{i})-D(p,\Lambda)-(\Lambda^{2}-M_{i}^{2})D(p,\Lambda)^{2})
\end{equation}

 With these preliminaries, we find the following expression for the pseudoscalar masses
 (squared), up to first order in the hairpin mass and (independently) in $L_5$ and $L_8$:

\begin{eqnarray}
\label{massformula}
 M^{2}_{ij} &=& \frac{\chi_{i}+\chi_{j}}{2}(1+\delta I_{ij})\times \nonumber \\
       &&\{1+\frac{1}{f^2}(8L_8 -4L_5)(\chi_{i}+\chi_{j})+\frac{1}{f^2}\delta(8L_8-4L_5)(2I_{ii}\chi_{i}
+2I_{jj}\chi_{j}+(\chi_{i}+\chi_{j})I_{ij}) \nonumber \\
 &+& 8\frac{1}{f^2}L_5 \delta J_{ij} \}  \\
 J_{ij} &\equiv& (I_{i}+I_{j}-(M^{2}_{ii}+M^{2}_{jj})I_{ij})/2
\end{eqnarray}

The quantities $\chi_{i}$ encode the quark masses: our data includes values for 9 different kappa
 values, so the indices $i,j$ above run from 1 to 9, allowing for 45 independent quark-antiquark
 combinations.  Thus 
\begin{equation}
\label{slopeparam}
    \chi_i =  2r_{0}m_{i}
\end{equation}
 where $r_{0}$ is a 
slope parameter (which we also extract from the fits) and we have used the 
pole value for the quark mass:

\begin{equation}
\label{qmass}
 m_{i} \equiv \ln(1+\frac{1}{2\kappa_{i}}-\frac{1}{2\kappa_{c}})
\end{equation}

\begin{figure}[p]
\renewcommand{\baselinestretch}{1.0}
\vspace*{-1.0cm}
\hspace*{-0.5cm}
\psfig{figure=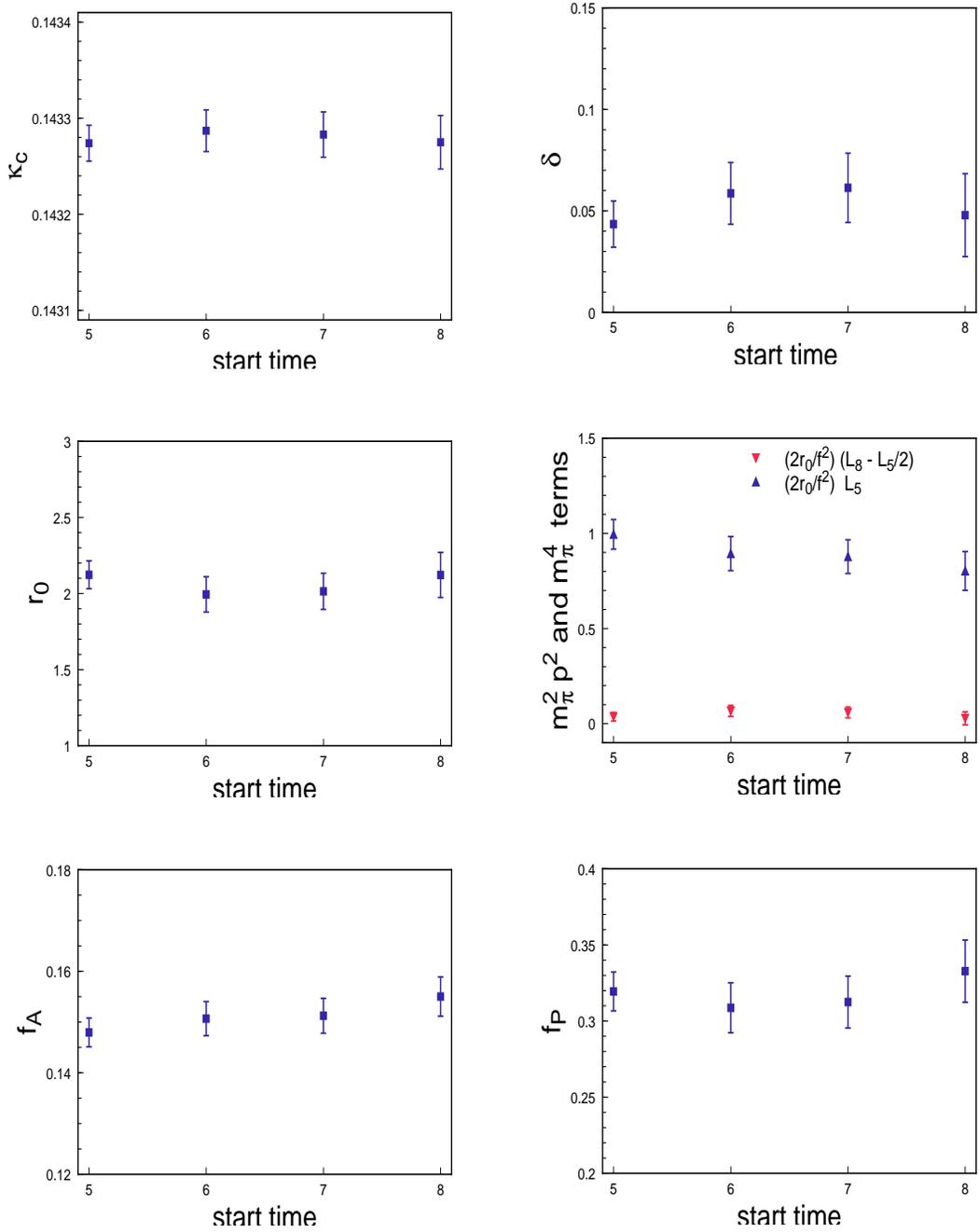,width=6.15in,height=7.5in}
\caption{Global Fits, using (\ref{massformula},\ref{fpformula},\ref{faformula}),
for $\kappa_{c},\delta,r_{0},L_{5,8},f_{A}=\sqrt{2}f,$and $f_{P}=r_{0}f_{A}$. 
Results are shown for time windows 5-11,6-11,7-11, and 8-11}
\label{fig:globfit}
\end{figure}

 Similar formulas were obtained for the pseudoscalar and axial decay constants $f_{P;ij},f_{A;ij}$ and
are listed in Appendix 2. We have performed global fits to the masses and decay constants for
 all 45 mesons in order
to extract the parameters $f,r_{0},\delta,L_{5}$ and $L_{8}$. The fits were performed for a
 variety of time-windows  (for the 12$^3$x24 b lattices, on time windows 5-11,6-11,7-11 and
 8-11) in order to isolate any remaining sensitivity to higher state 
contamination, and the results for the various chiral parameters, as a function of the initial
 time for the window, are shown in Fig[\ref{fig:globfit}]. With only 300 independent configurations, it is
 not possible to obtain a sufficiently stable covariance matrix to fit all 135 masses and decay
 constants, so these results reflect an uncorrelated fit to all meson parameters using (\ref{massformula}),
(\ref{fpformula}) and (\ref{faformula}).
  
\renewcommand{\baselinestretch}{1.7}

  To summarize our results, a global fit to all pseudoscalar masses and
 decay constants, using a time window of 6-11, gives a final value for the chiral log parameter of
\begin{equation}
 \delta= 0.059 \pm 0.015
\end{equation}
while the slope and critical kappa parameters in (\ref{slopeparam},\ref{qmass})
are determined as
\begin{equation} 
\label{r0result}
  r_{0} = 1.99 \pm 0.12
\end{equation}
\begin{equation}
\label{kcresult}
  \kappa_{c} = 0.143287 \pm 0.000022 
\end{equation} 
For the chiral breaking parameters $L_5$ and $L_8$, our fits give
\begin{equation}
   L_{5} = (2.5 \pm 0.5)\times 10^{-3}
\end{equation}
and                        
\begin{equation}
  L_{8}-\frac{1}{2}L_{5} = (0.2 \pm 0.1)\times 10^{-3}
\end{equation}
   The dimensionless chiral parameters $L_5,L_8$ are only roughly determined
 by phenomenology. Recent estimates \cite{chirpertth} give
\begin{equation}
   L_{5}(M_{\rho}) = (1.4 \pm 0.5)\times 10^{-3}
\end{equation}
renormalized at the rho mass, with the combination $L_{8}-\frac{1}{2}L_{5}$
consistent with zero:
\begin{equation}
  L_{8}-\frac{1}{2}L_{5} = (0.2 \pm 0.4)\times 10^{-3}
\end{equation}
 Finally, our result for the axial decay constant corresponds to
 a value for the pion decay constant (bare, in lattice units) of
\begin{equation}
  f = f_{\pi} = 0.1066 \pm 0.0024
\end{equation}

\newpage
\section{Summary and discussion}

The calculations presented above confirm all the essential features 
of anomalous quenched chiral behavior suggested by continuum calculations.
Our lattice studies lead to the following basic conclusions:
\begin{enumerate}
\item The MQA pole-shifting technique allows for accurate computation of
 meson and hairpin correlators down to small quark masses ($M_{\pi}<$ 200 MeV).
 Probing this region is essential in order to obtain reliable signatures
 of anomalous chiral behavior.
\item The hairpin vertex has only a small coupling to excited states, and
 very gentle momentum dependence. This suggests that it may be accurately
 modelled by a local mass-insertion term in a chiral Lagrangian.
\item Determination of the chiral log parameter $\delta$ using five separate
 methods gives consistent results. A summary of our results for this 
 parameter indicating the various methods used is displayed in Table 3.
 The overall average of our $C_{sw}=1.57$ methods gives $\delta = 0.065 \pm 0.013$.
 Our values for $\delta$ at $\beta=$ 5.7 are considerably smaller than those
 expected from a naive continuum analysis, but are in agreement with previous
 lattice estimates.
\item Using the MQA technique, meson properties (masses and decay constants)
 can be extracted with sufficient accuracy to allow a fit of higher order
 chiral parameters, such as $L_{5}$ and $L_{8}$.
\end{enumerate}

\begin{table}
\centering
\caption{Summary of results for chiral log parameter $\delta$}
\label{tab:deltares}
\vspace*{.2cm}
\begin{tabular}{|c|c|c|}
\hline 
Method              & $C_{sw}$        & $\delta$     \\
\hline
Witten-Veneziano    & 1.57            & 0.063(6)     \\
\hline
hairpin vertex      & 1.57            & 0.062(7)     \\
\hline
diagonal mesons     & 1.57            & 0.073(20)    \\
\hline
ratio fit, masses   & 1.57            & 0.060(16)    \\
ratio fit, $f_{P}/f_{A}$ & 1.57       & 0.071(10)    \\
\hline
global fit          & 1.57            & 0.059(15)     \\         
\hline
Witten-Veneziano    & 0               & 0.074(9)     \\
\hline
hairpin vertex      & 0               & 0.044(5)     \\
\hline
diagonal mesons     & 0               & 0.054(20)    \\
\hline
\end{tabular}
\end{table}

Careful quantitative studies of chiral behavior in quenched QCD in comparison to
quenched chiral perturbation theory can provide a great deal of insight into the 
connection between QCD and the effective chiral Lagrangian that describes its 
long range behavior in the limit of small quark mass.
Even if the numerical simulation of full QCD were not so expensive computationally, 
the study of chiral behavior in quenched QCD would still be of theoretical interest. 
For example, the Witten-Veneziano relation connects the
$\eta'$ mass to the topological susceptibility of {\it quenched} QCD. 
The geometric summation of multiple $\eta'$ mass insertions is only the simplest example
of how, in some cases, the most important effects of the full QCD 
fermion determinant can be incorporated into a
quenched result, with the guidance of chiral perturbation theory. Because of the
smallness of the parameter $\delta$, QCL effects are adequately described by one-loop
$\chi$PT, even for quark masses close to the physical up and down mass. It is
thus straightforward to apply appropriate and calculable QCL corrections to
quenched results. Of course the masses, decay constants, and higher order chiral
Lagrangian coefficients obtained in quenched
QCD will differ somewhat from those of the full theory, but all of the disturbing
structural properties of the quenched theory (lack of unitarity, absence of 
topological screening, etc.) can be systematically repaired in
the context of chiral perturbation theory. Further precision studies of anomalous
chiral behavior in the quenched meson and baryon spectrum should provide additional
insight into the origin and structure of chiral symmetry in QCD. The results presented
in this paper provide strong support for the usefulness of the MQA technique to
facilitate these studies.

Ideally, similar studies should be performed using an exactly chirally
symmetric Dirac operator which satisfies Ginsparg-Wilson relations, e.g. the
Neuberger operator\cite{Neuberger}.  Explicit Ginsparg-Wilson chiral symmetry would 
resolve the exceptional configuration problem {\it ab initio}. Unfortunately,
such operators are necessarily not ultralocal\cite{Horvath}, and are difficult
to invert or diagonalize numerically. 

The MQA method attempts to account for the most salutary effect
of an explicitly chirally symmetric approach. 
The underlying assumption in this procedure is that the most important effect of the
explicit chiral symmetry breaking contained in the Wilson-Dirac operator is the 
real displacement of its small eigenvalues. 
The MQA method merely removes these displacements in a compensated manner.
The principal disadvantage of the method is its apparent lack of locality.
Using a basis of hopping terms (nearest-neighbor, next-nearest neighbor, etc.),
it may be possible to identify terms in an ultralocal expansion of the
Dirac operator which correspond most closely to an MQA improved 
Wilson-Dirac operator.
The additional hopping terms would have the effect of reducing 
the dispersion of the small real eigenvalues as well as inducing small
rotations of the basis wavefunctions, etc.
Such an analysis has not been carried out yet. 
The clear quantitative success of the MQA procedure in restoring desired
chiral behavior is a promising indication that such eigenmode-based methods
can be both efficient and effective in removing the dominant spurious 
effects of chiral symmetry breaking contained in the Wilson-Dirac formulation
of lattice fermions.

\appendix
\newpage
\section*{Appendix 1: The allsource method}

The allsource method, first applied to the $\eta'$ mass calculation in Ref.
\cite{Kuramashi}, is used here to calculate hairpin diagrams as well as the
pseudoscalar charge for the determination of winding numbers. We have also 
introduced a generalization of this allsource method which allows the calculation 
of closed loops which originate from a smeared source, i.e. the 
two ends of the quark propagator are contracted over color and  spin, but are
spatially separated with an exponential weight function. The method relies
on the fact that gauge noninvariant terms in the calculation will cancel out due
to random SU(3) gauge phases. For example, in the calculation of a single quark
loop, the closed loop terms where the 
quark starts and ends on the same point
(or on the same exponential source in the smeared calculation) add coherently
when summed over sites, while loops which start and end on different sources have random phases and cancel. 

\begin{figure}
\vspace*{0.5cm}
\includegraphics{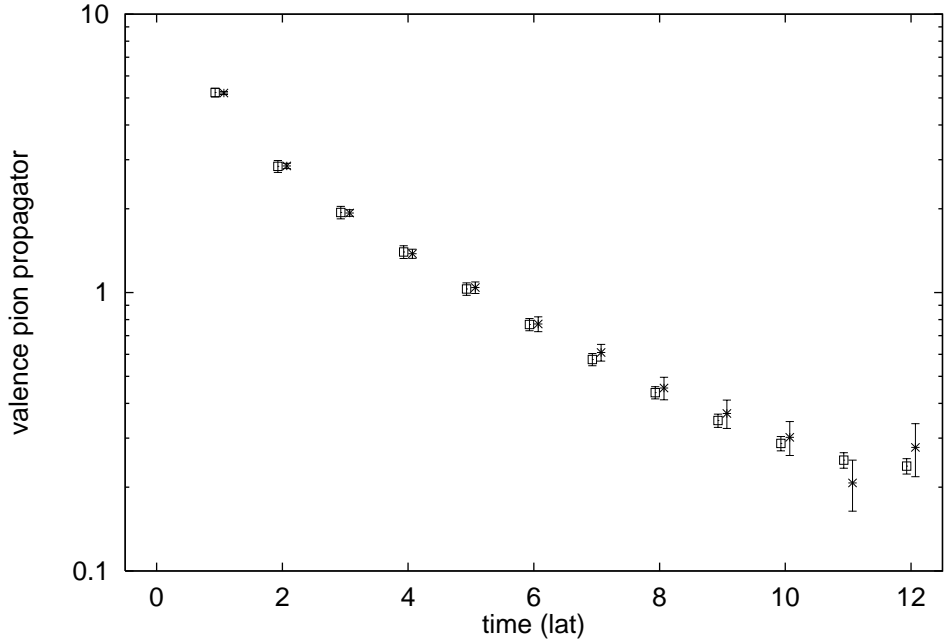}
\vspace{6.5cm}
\caption[]{Comparison of valence pion propagator computed by allsource method ($\times$'s) and
conventional method (boxes). Data points are offset slightly for clarity. }
\label{fig:allsource}
\end{figure}

While the cancellation of random gauge phases in the allsource method should
work arbitrarily well for a large enough ensemble of gauge configurations, it
is very instructive to test this method in a situation where we know the
exact gauge invariant answer which can be used to check the accuracy of the
random phase cancellation. To carry out such a comparison, we have calculated
the ordinary valence pion propagator using the allsource technique. This
calculation uses the same allsource propagators that were used to calculate
hairpin correlators. In the hairpin calculation, we computed the correlator 
of two closed loops by contracting the two color indices of each allsource
propagator with a fixed time separation between the two propagators. 
To calculate the valence propagator, we instead take the same two allsource 
propagators with fixed time separation and cross-contract the color indices
to form a single loop from the two propagators and then project out the color singlet component.
(The last step amounts to leaving out the terms in which all four color indices
are equal and then multiplying by a factor of $3/2$.)
In Fig[\ref{fig:allsource}], the valence pion propagator calculated by the allsource method 
is compared with the results of the standard calculation using local and smeared 
sources on a fixed timeslice. The results are shown for hopping parameter $\kappa=.1425$, 
but similar agreement is obtained at all kappa values. 

The agreement is excellent and well within statistical errors. For small times
($t<5$), the errors on the allsource calculation are actually smaller than those
from a fixed source. (Remember that the allsource calculation allows the meson
propagator to be averaged over all locations on the lattice, thus increasing
the effective statistics relative to the fixed source method.)
Unlike the fixed 
source calculation however, the statistical errors for the allsource calculation are 
more-or-less constant in absolute magnitude for all time separations. This results 
in a signal-to-noise ratio that gets rapidly worse as we go out in time, just as in the 
hairpin calculation. This roughly constant noise level is presumably the effect of 
incomplete cancellation of random gauge phases. Because of this, the allsource 
method is an inferior way of studying the asymptotic behavior of the pion 
propagator. Nevertheless, for short times it accurately reproduces the results of the 
standard method. It is a fortuitous circumstance that the hairpin correlator 
is found to be almost entirely free of excited state contamination. This allows
us to extract the ground state vertex insertion from its value at relatively short times where the allsource technique is accurate.

\newpage
\section*{Appendix 2: Quenched chiral results for decay \linebreak constants}

 A next to leading order chiral perturbation theory calculation of the
pseudoscalar and axial decay constants can be carried out along the
same lines as discussed in Section 8 for the pseudoscalar mass
spectrum.
Starting from the Lagrangian (\ref{eq:chilag}) and using 
the notations introduced there, we find
\begin{eqnarray}
\label{fpformula}
   f_{P;ij} &=& \sqrt{2}fr_{0}(1+\frac{1}{4}\delta(I_{ii}+I_{jj}+2I_{ij}))\times \nonumber \\
         &\{& 1+\frac{1}{f^2}(8L_8 -2L_5)(\chi_{i}+\chi_{j})(1+\delta(2I_{ii}\chi_{i}+2I_{jj}\chi_{j}
 +(\chi_{i}+\chi_{j})I_{ij})/(\chi_i + \chi_j))\hspace*{1.5em} \nonumber \\  
  &-&\frac{2}{f^2}L_{5}\delta(I_{ii}\chi_{i}+I_{jj}\chi_{j}+(\chi_{i}+\chi_{j})I_{ij}) \nonumber \\
  &+&\frac{4}{f^2}L_{5}\delta(J_{ii}+J_{jj}) \}
\end{eqnarray}
for the pseudoscalar decay constant, and
\begin{eqnarray}
\label{faformula}
   f_{A;ij} &=& \sqrt{2}f(1+\frac{1}{4}\delta(I_{ii}+I_{jj}-2I_{ij}))\times \nonumber \\
     &\{& 1 + \frac{2}{f^2}L_{5}(\chi_{i}+\chi_{j}) 
     + 2\frac{1}{f^2}L_{5}\delta(2(\chi_{i}+\chi_{j})I_{ij}-\chi_{i}I_{ii}-\chi_{j}I_{jj}) \} \hspace*{8.5em}
\end{eqnarray}
for the axial decay constant. 
These formulas, together with  
\begin{eqnarray}
\label{massformulaB}
 M^{2}_{ij} &=& \frac{\chi_{i}+\chi_{j}}{2}(1+\delta I_{ij})\times \nonumber \\
       &&\{1+\frac{1}{f^2}(8L_8 -4L_5)(\chi_{i}+\chi_{j})+\frac{1}{f^2}\delta(8L_8-4L_5)(2I_{ii}\chi_{i}
+2I_{jj}\chi_{j}+(\chi_{i}+\chi_{j})I_{ij}) \nonumber \\
 &+& 8\frac{1}{f^2}L_5 \delta J_{ij} \}   
\end{eqnarray}
for the pseudoscalar masses,
were used to perform the global fits described in Section 8.

\newpage 
{\noindent \Large \bf Acknowledgements}

  The work of W. Bardeen and E. Eichten was performed at the Fermi National
Accelerator Laboratory, which is operated by University Research Association,
Inc., under contract DE-AC02-76CHO3000. 
The work of A. Duncan was supported in part by NSF grant PHY97-22097.
The work of H. Thacker was supported in part by the Department of Energy
under grant DE-FG02-97ER41027.

\end{document}